\documentclass[11pt, notitlepage]{article} 
\usepackage{a4}
\usepackage{epstopdf}
\usepackage{xcolor}

\definecolor{TUMblue}{RGB}{0, 101, 189}
\definecolor{TUMlightblue}{RGB}{100,160,200}
\definecolor{TUMgreen}{RGB}{162,173,0}
\definecolor{TUMorange}{RGB}{227,114,034}
\definecolor{TUMivory}{RGB}{218,215,203}
\definecolor{fgreen}{RGB}{34,139,34}

\usepackage{natbib}

\usepackage{hyperref}
\hypersetup{
	colorlinks=true,
	linkcolor=TUMblue,
	citecolor=TUMblue,
	filecolor=TUMblue,
	urlcolor=TUMblue
}

\usepackage{etoolbox}
\makeatletter

\pretocmd{\NAT@citex}{%
	\let\NAT@hyper@\NAT@hyper@citex
	\def\NAT@postnote{#2}%
	\setcounter{NAT@total@cites}{0}%
	\setcounter{NAT@count@cites}{0}%
	\forcsvlist{\stepcounter{NAT@total@cites}\@gobble}{#3}}{}{}
\newcounter{NAT@total@cites}
\newcounter{NAT@count@cites}
\def\NAT@postnote{}

\def\NAT@hyper@citex#1{%
	\stepcounter{NAT@count@cites}%
	\hyper@natlinkstart{\@citeb\@extra@b@citeb}#1%
	\ifnumequal{\value{NAT@count@cites}}{\value{NAT@total@cites}}
	{\ifNAT@swa\else\if*\NAT@postnote*\else%
		\NAT@cmt\NAT@postnote\global\def\NAT@postnote{}\fi\fi}{}%
	\ifNAT@swa\else\if\relax\NAT@date\relax
	\else\NAT@@close\global\let\NAT@nm\@empty\fi\fi
	\hyper@natlinkend}
\renewcommand\hyper@natlinkbreak[2]{#1}

\patchcmd{\NAT@citex}
{\ifNAT@swa\else\if*#2*\else\NAT@cmt#2\fi
	\if\relax\NAT@date\relax\else\NAT@@close\fi\fi}{}{}{}
\patchcmd{\NAT@citex}
{\if\relax\NAT@date\relax\NAT@def@citea\else\NAT@def@citea@close\fi}
{\if\relax\NAT@date\relax\NAT@def@citea\else\NAT@def@citea@space\fi}{}{}

\makeatother

\usepackage{courier}
\usepackage{amssymb, graphicx}
\usepackage{subfigure}
\usepackage{amsmath}
\usepackage{amsthm}
\usepackage{verbatim}
\usepackage{dsfont}
\usepackage{geometry}
\usepackage{pdflscape}
\usepackage{multirow}
\usepackage{aliascnt}
\usepackage{verbatim}
\usepackage{graphicx}
\usepackage{float}
\usepackage{tikz}
\usetikzlibrary{calc,shapes,arrows,decorations.pathmorphing,graphs,positioning,backgrounds,arrows.meta}
\usepackage{latexsym}
\usepackage{mathtools}
\usepackage{mathrsfs}
\usepackage{bm}
\usepackage{bbm}
\usepackage{shadethm}
\usepackage{enumerate}
\usepackage{colortbl}
\usepackage{framed}
\colorlet{shadecolor}{gray!25}
\usepackage{booktabs}
\usepackage{longtable}
\usepackage{multirow}
\usepackage{rotating}
\usepackage{chngpage}
\usepackage{appendix}
\usepackage{pdfpages}
\usepackage{adjustbox}

\usepackage{algorithm}
\usepackage[noend]{algpseudocode}
\algnewcommand{\IIf}[1]{\State\algorithmicif\ #1\ \algorithmicthen}
\algnewcommand{\EndIIf}{\unskip\ \algorithmicend\ \algorithmicif}
\makeatletter
\def\BState{\State\hskip-\ALG@thistlm}
\makeatother

\newcommand{\mynewtheorem}[2]{
	\newaliascnt{#1}{dummy}
	\newtheorem{#1}[#1]{#2}
	\aliascntresetthe{#1}
	\expandafter\def\csname #1autorefname\endcsname{#2}
}

\bibpunct[\textcolor{TUMblue}{, }]{\textcolor{TUMblue}{(}}{\textcolor{TUMblue}{)}}{\textcolor{TUMblue}{;}}{\textcolor{TUMblue}{a}}{\textcolor{TUMblue}{}}{\textcolor{TUMblue}{,}}
\makeatletter
\renewcommand\eqref[1]{%
	\textup{\color{TUMblue}\tagform@{\ref{#1}}}%
}

\providecommand{\keywords}[1]{\vspace{0.25cm} \noindent \textit{Keywords:} #1}

\theoremstyle{definition}
\mynewtheorem{thm}{Theorem}
\mynewtheorem{defi}{Definition}
\mynewtheorem{lem}{Lemma}
\mynewtheorem{cor}{Corollary}
\mynewtheorem{proposition}{Proposition}
\mynewtheorem{example}{Example}
\mynewtheorem{alg}{Algorithm}
\mynewtheorem{remark}{Remark}

\DeclareMathOperator*{\argmax}{arg\,max}

\definecolor{mygray}{gray}{0.85}
\def\bm#1{\mbox{\boldmath $#1$}}

\def\bms#1{\boldsymbol{#1}} 

\usetikzlibrary{shapes,arrows,decorations.pathmorphing,graphs,positioning,backgrounds}

\tikzstyle{VineNode} = [ellipse, fill = white, draw = black, text = black, align = center, minimum height = 1cm, minimum width = 1cm]
\tikzstyle{DummyNode}  = [draw = none, fill = none, text = black]
\tikzstyle{TreeLabels} = [draw = none, fill = none, text = black] 
\newcommand{\xshiftNodes}{0.7*\linewidth}
\newcommand{\yshiftLabels}{-.25cm}

\usepackage{etoolbox}
\makeatletter
\patchcmd{\hyper@makecurrent}{%
	\ifx\Hy@param\Hy@chapterstring
	\let\Hy@param\Hy@chapapp
	\fi
}{%
	\iftoggle{inappendix}{
		\@checkappendixparam{chapter}%
		\@checkappendixparam{section}%
		\@checkappendixparam{subsection}%
		\@checkappendixparam{subsubsection}%
		\@checkappendixparam{paragraph}%
		\@checkappendixparam{subparagraph}%
	}{}%
}{}{\errmessage{failed to patch}}

\newcommand*{\@checkappendixparam}[1]{%
	\def\@checkappendixparamtmp{#1}%
	\ifx\Hy@param\@checkappendixparamtmp
	\let\Hy@param\Hy@appendixstring
	\fi
}
\makeatletter

\newtoggle{inappendix}
\togglefalse{inappendix}

\apptocmd{\appendix}{\toggletrue{inappendix}}{}{\errmessage{failed to patch}}
\apptocmd{\subappendices}{\toggletrue{inappendix}}{}{\errmessage{failed to patch}}

\usepackage{sectsty}
\allsectionsfont{\sffamily}

\usepackage{titlesec}
\titleformat{\chapter}[display]
{\normalfont\sffamily\LARGE\bfseries\centering}
{\chaptertitlename\ \thechapter}{20pt}{\LARGE}

\usepackage{caption}
\usepackage{footnote}

\begin{document}
	
	{	\renewcommand*{\thefootnote}{\fnsymbol{footnote}}
		\title{\textbf{\sffamily Nonparametric C- and D-vine based quantile regression}}
		
		\date{\small \today}
		\newcounter{savecntr1}
		\newcounter{restorecntr1}
		\newcounter{savecntr2}
		\newcounter{restorecntr2}

		
		\author{Marija Tepegjozova\setcounter{savecntr1}{\value{footnote}}\thanks{Department of Mathematics, Technische Universit{\"a}t M{\"u}nchen, Boltzmannstra{\ss}e 3, 85748 Garching, Germany (email: \href{mailto:m.tepegjozova@tum.de}{m.tepegjozova@tum.de} (corresponding author),  \href{mailto:cczado@ma.tum.de}{cczado@ma.tum.de})},
			Jing
			Zhou\setcounter{restorecntr1}{\value{footnote}}\thanks{ORStat and Leuven Statistics Research Center, KU Leuven, Naamsestraat 69-box 3555 Leuven, Belgium (email: \href{mailto:jing.zhou@kuleuven.be}{jing.zhou@kuleuven.be}, \href{mailto:gerda.claeskens@kuleuven.be}{gerda.claeskens@kuleuven.be})},
			Gerda Claeskens$^{\dagger}$
			and Claudia Czado$^{*}$
		}
		\maketitle
		
	}

	\begin{abstract}
	{ Quantile regression is a field with steadily growing importance in statistical modeling. It is a complementary method to linear regression, since computing a range of conditional quantile functions provides more accurate modeling of the stochastic relationship among variables, especially in the tails. We introduce a   non-restrictive and highly flexible nonparametric quantile regression approach based on C- and D-vine copulas.
		Vine copulas allow for separate modeling of marginal distributions and the dependence structure in the data, and can be expressed through a
		graphical structure consisting of a sequence of linked trees.
		This way, we obtain a quantile regression model that overcomes  typical issues
		of quantile regression such as quantile crossings or collinearity, the need for transformations and interactions of variables.
		Our approach incorporates a  two-step ahead ordering of variables, by
		maximizing the conditional log-likelihood of the tree sequence, while taking into account the next two tree levels.
		We show that the nonparametric conditional quantile estimator is consistent. The performance of the proposed methods is evaluated in both low- and high-dimensional settings using simulated and real-world data. The results support the superior prediction ability of the proposed models.}
\end{abstract}


\maketitle

\section{Introduction}\label{introduction1}
As a robust alternative to the ordinary least squares regression, which estimates the conditional mean, quantile regression \citep{koenker1978regression} focuses on the conditional quantiles. This method has been studied extensively in statistics, economics, and finance.
The pioneer literature by \citet{Koenker2005} investigated linear quantile regression systematically. It presented properties of the estimators including asymptotic normality and consistency, under various assumptions such as independence of the observations, independent and identically distributed (i.i.d.) errors with continuous distribution, and predictors having bounded second moment. Subsequent extensions of linear quantile regression have been intensively studied, see for example adapting quantile regression in the Bayesian framework \citep{yu2001bayesian}, for longitudinal data \citep{koenker2004quantile}, time-series models \citep{xiao2009conditional}, high-dimensional models with $l_1$-regularizer \citep{belloni2011etal}, nonparametric estimation by kernel weighted local linear fitting \citep{yu1998local}, and by additive models \citep{koenker2011additive, fenske2011identifying}, etc.
The theoretical analysis of the above-mentioned extensions is based on imposing additional assumptions such as samples that are i.i.d. (see for example \citet{yu1998local,belloni2011etal}), or that are generated by a known additive function (see for example \citet{koenker2011additive, koenker2004quantile}). Such assumptions, which guarantee the performance of the proposed methods for certain data structures, cause concerns in applications due to the uncertainty of the real-world data structures.
\citet{bernard2015conditional} addressed other potential concerns such as quantile crossings and model-misspecification, when the dependence structure of the response variables and the predictors does not follow a Gaussian copula.
Flexible models without assuming  homoscedasticity, or a linear relationship between the response and the predictors are of interest. Recent research on dealing with this issue includes quantile forests \citep{meinshausen2006quantile, li2017forest,athey2019generalized} inspired by the earlier work of random forests \citep{breiman2001random} and modeling conditional quantiles using copulas (see also \citet{noh2013copula, noh2015semiparametric, chen2009copula}).

\noindent	Vine copulas in the context of conditional quantile prediction have been investigated by \citet{kraus2017d} using drawable vine copulas (D-vines), \citet{chang2019prediction} and most recently, \citet{zhu2021simplified} using restricted regular vines (R-vines).  The approach of \citet{chang2019prediction} is based on first finding the locally optimal regular vine structure among all predictors and then adding the response to each selected tree in the vine structure as a leaf, as also followed by \citet{bauer2016pair} in the context of non-Gaussian conditional independence testing. The procedure in \citet{chang2019prediction} allows for a recursive determination of the response quantiles, which is restricted through the prespecified dependence structure among predictors. The latter might not be the one maximizing the conditional response likelihood,  which is the main focus in regression setup.
The approach of \citet{kraus2017d} is based on optimizing the conditional log-likelihood and selecting predictors sequentially until no improvement of the conditional log-likelihood is achieved. This approach based on the conditional response likelihood is more appropriate to determine the associated response quantiles. 	
Further, the intensive simulation study in \citet{kraus2017d}  showed the superior performance of the D-vine copulas based quantile regression compared to various quantile regression methods, i.e., linear quantile regression \citep{koenker1978regression}, boosting additive quantile regression \citep{Koenker2005, koenker2011additive, fenske2011identifying}, nonparametric quantile regression \citep{li2013optimal}, and semiparametric quantile regression \citep{noh2015semiparametric}.In parallel to our work, \citet{zhu2021simplified} proposed an extension of this D-vine based forward regression  to a restricted R-vine forward regression with comparable performance to the D-vine regression.  Thus, the D-vine quantile regression will be our benchmark model.\\
We extend the method of \citet{kraus2017d} in two ways: (1) our approach is applicable to both C-vine and D-vine copulas; (2) a  two-step ahead construction is introduced, instead of the one-step ahead construction. Since the two-step ahead construction is the main difference between our method and \citet{kraus2017d}, we further explain the second point in more detail. Our proposed method proceeds by adding predictors to the model sequentially. However, in contrast to \citet{kraus2017d} with only one variable ahead, our new approach proposes to look up two variables ahead for selecting the variable to be added in each step. The general idea of this  two-step ahead algorithm is as follows: in each step of the algorithm, we study combinations of two variables to find the variable, which in combination with the other improves the conditional log-likelihood the most. Thus, in combination with a forward selection method, this two-step ahead algorithm allows us to construct nonparametric quantile estimators that improve the conditional log-likelihood in each step and, most importantly, take possible future improvements into account.
Our method is applicable to both low and  high-dimensional data.
By construction, quantile crossings are avoided. All marginal densities and copulas are estimated nonparametrically, allowing more flexibility than parametric specifications. \citet{kraus2017d} addressed the necessity and possible benefit of the nonparametric estimation of bivariate copulas in the quantile regression framework. This construction permits a large variety of dependence structures, resulting in a well-performing conditional quantile estimator.  Moreover, extending to the C-vine copula class, in addition to the D-vine copulas, provides greater flexibility.

\noindent The paper is organized as follows. Section~\ref{section:background} introduces the general setup, the concept of C-vine and D-vine copulas and the nonparametric approach for estimating copula densities. Section~\ref{section:vinebasedintro} describes the vine based approach for quantile regression.
The new two-step ahead forward selection algorithms are described in Section~\ref{section:main}. We investigate in Proposition~\ref{thm:consistency} the consistency of the conditional quantile estimator for given variable orders. The finite sample performance of the vine based conditional quantile estimator is evaluated in Section~\ref{section:simulation} by several quantile related measurements in various simulation settings. We apply the newly introduced algorithms to low- and high-dimensional real data in Section~\ref{section: real data}. In Section~\ref{section:disscussion} we conclude and discuss  possible directions of future research.
\newpage
\section{Theoretical background}\label{section:background}
Consider the random vector  $\bm{X} = (X_1, \ldots, X_d)^T$ with observed values $\bm{x}=(x_1, \ldots, x_d)^T$, joint distribution and density function $F$ and $f$, marginal distribution and density functions $F_{X_j}$ and $f_{X_j}$ for $X_j, j=1,\ldots, d$.
Sklar's theorem \citep{sklar1959fonctions} allows to represent any multivariate distribution in terms of its marginals $F_{X_j}$ and a copula $C$ encoding the dependence
structure. In the continuous case, $C$ is unique and satisfies
$F(\bm{x})  =   C(F_{X_1}(x_1), \ldots, F_{X_d}(x_d))$ and $f(\bm{x})  =  c(F_{X_1}(x_1), \ldots, F_{X_d}(x_d))[\prod_{j=1}^d f_{X_j}(x_j)]$,
where $c$ is the density function of the copula $C$.
To characterize the dependence structure of $\bm{X}$, we transform each $X_j$ to a uniform variable $U_j$ by applying the probability integral transform, i.e. $U_j \coloneqq F_{X_j}(X_j), \; j = 1, \ldots, d$. Then the random vector $\bm U = (U_1, \ldots, U_d)^T$ with observed values
$(u_1, \ldots, u_d)^T$ has a copula as a joint distribution denoted as $C_{U_1,\ldots, U_d}$ with associated copula density function $c_{U_1, \ldots, U_d}$.
While the catalogue of bivariate parametric copula families is large, this is not true
for $d > 2$. Therefore conditioning was applied to construct multivariate copulas using only bivariate copulas as building blocks. \citet{joe1996families} gave the
first pair copula construction for $d$ dimensions in terms of distribution functions,
while \citet{bedford2002vines} independently developed constructions in terms
of densities together with a graphical building plan, called a regular vine tree structure. It consists of a set of linked trees $T_1, \ldots, T_d$ (edges in tree $T_j$ become nodes in tree $T_{j+1}$) satisfying a proximity condition, which allows to identify all possible
constructions. Each edge of the trees is associated with a pair copula $C_{U_i, U_{j}; \bms{U}_D}$, where $D$ is a subset of indices not containing $i,j$. In this case the set $\{i,j\}$ is called the conditioned set, while $D$ is the conditioning set.  A joint density using the class of vine copulas is then the product of all pair copulas identified by the tree structure evaluated at appropriate conditional distribution functions $F_{X_{j}|\bms{X}_D}$ and the product of the marginal densities $f_{X_j},j=1,\ldots,d$.
A detailed  treatment of vine copulas together with estimation methods and model choice approaches are given, for example in \citet{joe2014dependence} and \citet{czado2019analyzing}.

\begin{figure}[!htb]
	\begin{minipage}{.2\textwidth}

		\resizebox{!}{3in}{	
			\begin{tikzpicture}	[every node/.style = VineNode, node distance =1.5cm]
				\node[TreeLabels] (T1)       {$T_1$}
				node             (1)         [right of = T1] {1}			
				node[DummyNode]  (Dummy1)         [right of = 1, xshift = \xshiftNodes] {}
				node             (3)         [above of = Dummy1] {3}
				node             (2)         [above of = 3] {2}
				node             (4)         [below of = 3] {4}
				;	    	
				\draw (1) to node[draw=none, fill = none, above, yshift = \yshiftLabels] {1,2} (2);
				\draw (1) to node[draw=none, fill = none, above, yshift = \yshiftLabels] {1,3} (3);
				\draw (1) to node[draw=none, fill = none, above, yshift = \yshiftLabels] {1,4} (4);
				
				\node (13)       [below of = 4 ]{1,3}
				node             (14)         [below of = 13] {1,4}	
				node             (12)         [left of = 14, xshift = -\xshiftNodes] {1,2}
				node[TreeLabels] (T2)       [left of = 12] {$T_2$}	
				;	    	
				\draw (12) to node[draw=none, fill = none, above, yshift = \yshiftLabels] {2,3;1} (13);
				\draw (12) to node[draw=none, fill = none, above, yshift = \yshiftLabels] {2,4;1} (14);
				\node[TreeLabels] (T3) [below of = T2 ]  {$T_3$}
				node (231)             [right of = T3] {2,3;1}
				node (241)             [right of = 231, xshift = \xshiftNodes] {2,4;1}
				;
				\draw (231) to node[draw=none, fill = none, above, yshift = \yshiftLabels] {3,4;1,2} (241);
				
				\node[DummyNode] (T4)   [below of = T3]    {}
				node             (1423)         [right of = T4, xshift = 1.8cm ] {3,4;1,2};
		\end{tikzpicture}	}	
		\centering
	\end{minipage}%
	\hspace{2.9cm}
	\begin{minipage}{0.5\textwidth}
		\raggedright
		\renewcommand{\xshiftNodes}{0.0075*\linewidth}
		\renewcommand{\yshiftLabels}{.0cm}
		\resizebox{3.5in}{!}{
			\begin{tikzpicture}[every node/.style = VineNode, node distance =1.5cm]
				\node[TreeLabels] (T1)       {$T_1$}
				node             (1)         [right of = T1] {1}			
				node[DummyNode]  (Dummy1)         [right of = 1, xshift = \xshiftNodes] {}
				node             (2)         [right of = Dummy1, xshift = \xshiftNodes] {2}			
				node[DummyNode]  (Dummy2)         [right of = 2, xshift = \xshiftNodes] {}
				node             (3)         [right of = Dummy2, xshift = \xshiftNodes] {3}			
				node[DummyNode]  (Dummy3)         [right of = 3, xshift = \xshiftNodes] {}
				node             (4)         [right of = Dummy3, xshift = \xshiftNodes] {4}			
				;	    	
				\draw (1) to node[draw=none, fill = none, above, yshift = \yshiftLabels] {1,2} (2);
				\draw (2) to node[draw=none, fill = none, above, yshift = \yshiftLabels] {2,3} (3);
				\draw (3) to node[draw=none, fill = none, above, yshift = \yshiftLabels] {3,4} (4);
				
				\node[TreeLabels] (T2)   [below of = T1]    {$T_2$}
				node             (12)         [below of = Dummy1] {1,2}
				node             (23)         [below of = Dummy2] {2,3}
				node             (34)         [below of = Dummy3] {3,4}
				node[DummyNode]             (Dummy5)         [below of = 2] {}
				node[DummyNode]             (Dummy6)         [below of = 3] {}
				node [DummyNode]            (Dummy7)         [below of = 4] {}					
				;	    	
				\draw (12) to node[draw=none, fill = none, above, yshift = \yshiftLabels] {1,3;2} (23);
				\draw (23) to node[draw=none, fill = none, above, yshift = \yshiftLabels] {2,4;3} (34);

				\node[TreeLabels] (T3)   [below of = T2]    {$T_3$}
				node             (132)         [below of = Dummy5] {1,3;2}
				node             (243)         [below of = Dummy6] {2,4;3}
				
				node[DummyNode]             (Dummy8)         [below of = 23] {}
				node[DummyNode]             (Dummy9)         [below of = 34] {}
				;	    	
				\draw (132) to node[draw=none, fill = none, above, yshift = \yshiftLabels] {1,4;2,3} (243);

				\node[TreeLabels] (T4)   [below of = T3]    {}
				node             (1423)         [below of = Dummy8] {1,4;2,3}
				;

		\end{tikzpicture}}
	\end{minipage}
	\caption{ C-vine tree sequence (left panel) and a D-vine tree sequence (right panel) in 4 dimensions.}
	\label{fig:examplecdvine}
\end{figure}
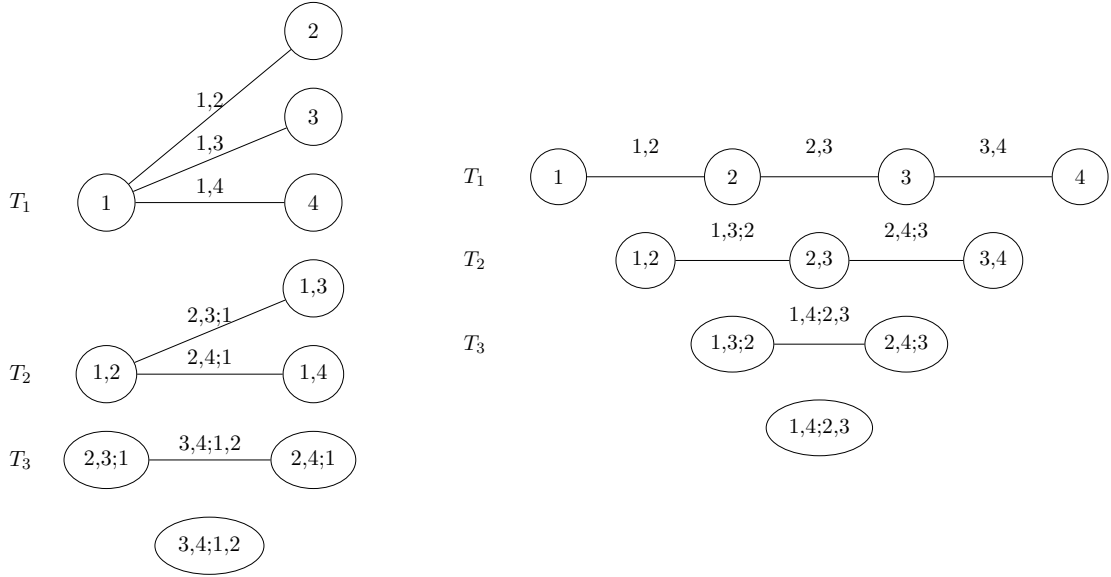

\noindent 	Since we are interested in simple copula based estimation methods for conditional quantiles, we
restrict to two subclasses of the regular vine tree structure, namely the D- and C-vine structure. We show later that these structures allow us to express conditional distribution and quantiles in closed form. In the D-vine tree structure all trees are paths, i.e. all nodes have degree $\leq 2$. Nodes with degree 1 are called leaf nodes.  A C-vine structure occurs, when all trees are stars with a root node in the
center. The right and left panel of Figure~\ref{fig:examplecdvine} illustrates  a D-vine and a C-vine tree sequence in four dimensions, respectively.

\noindent For these sub classes we can easily give the corresponding vine density \citep[Chapter~4]{czado2019analyzing}. For a D-vine density we have a permutation $s_1,\ldots ,s_d$ of $1,\ldots ,d$ such that
\begin{equation}
	\label{eq:d-vine}
	\begin{aligned}
		f( x_1,\hdots,x_d)= &  \prod_{j=1} ^{d-1}\prod_{i=1}^{d-j} c_{U_{s_i},U_{s_{i+j}};U_{s_{i+1}},\hdots,U_{s_{i+j-1}}} \left(  F_{X_{s_i}|X_{s_{i+1}},\hdots,X_{s_{i+j-1}}} (x_{s_i}|x_{s_{i+1}}, \hdots, x_{s_{i+j-1}}) , \right.  \\
		& \left.  F_{X_{s_{i+j}}|X_{s_{i+1}},\hdots,X_{s_{i+j-1}}} (x_{s_{i+j}}|x_{s_{i+1}}, \hdots, x_{s_{i+j-1}})	
		\right)   \cdot    \prod_{k=1}^{d}f_{X_{s_k}} (x_{s_k}),
	\end{aligned}
\end{equation}
while for a C-vine density the following representation
holds
\begin{equation}
	\label{eq:c-vine}
	\begin{aligned}
		f( x_1,\hdots,x_d)= & \prod_{j=1} ^{d-1}\prod_{i=1}^{d-j} c_{U_{s_j},U_{s_{j+i}};U_{s_1},\hdots,U_{s_{j-1}}}
		\left( F_{X_{s_j}|X_{s_1},\hdots,X_{s_{j-1}}} (x_{s_j}|x_{s_1}, \hdots, x_{s_{j-1}}),  \right. \\ & \left.  F_{X_{s_{j+i}}|X_{s_1},\hdots,X_{s_{j-1}}}  (x_{s_{j+i}}|x_{s_1}, \hdots, x_{s_{j-1}})  \right)  \cdot   \prod_{k=1}^{d}f_{X_{s_k}} (x_{s_k}).
	\end{aligned}
\end{equation}
To determine the needed conditional distribution
$F_{X_{j}|\bms{X}_D}$
in \eqref{eq:d-vine} and \eqref{eq:c-vine} for appropriate choices of $j$ and $D$, the recursion discussed in \citet{joe1996families} is available. Using  $u_j=F_{X_{j}|\bms{X}_D}(x_j|\bms{x}_D)$ for $j=1,\ldots,d$ we can express them as compositions of h-functions. These are defined in general as  $h_{U_i|U_j;\bms{U}_D}(u_i|u_j; \bm{u}_D) = \frac{\partial}{\partial u_j}C_{U_i,U_j;\bms{U}_D}(u_i,u_j;\bm{u}_D)$.
Additionally we made in
\eqref{eq:d-vine} and \eqref{eq:c-vine} the simplifying assumption \citep[Section 5.4]{czado2019analyzing}, that is, the copula function
$C_{U_i,U_j;\bms{U}_D}$ does not depend on the specific conditioning value of $\bm{u}_D$, i.e. $C_{U_i,U_j;\bms{U}_D}(u_i,u_j;\bm{u}_D)=C_{U_i,U_j;\bms{U}_D}(u_i,u_j)$. The dependence on $\bm{u}_D$ in \eqref{eq:d-vine} and \eqref{eq:c-vine} is completely captured by the arguments of the pair copulas. This assumption is often made for tractability reasons in higher dimensions (\citet{haff2010simplified} and \citet{stoeber2013simplified}). It implies further, that the h-function satisfies  $h_{U_i|U_j;\bms{U}_D}(u_i|u_j; \bm{u}_D) = \frac{\partial}{\partial u_j}C_{U_i,U_j;\bms{U}_D}(u_i,u_j)=C_{U_i|U_j;\bms{U}_D}(u_i|u_j)$  and is independent of $\bm{u}_D$.

\subsection{Nonparametric estimators of the copula densities and h-functions}\label{sectionnonpar}


There are many methods to estimate a bivariate copula density $c_{U_i, U_j}$ nonparametrically. Examples are the transformation estimator \citep{charpentier2007estimation}, the transformation local likelihood estimator \citep{geenens2017probit}, the tapered transformation estimator \citep{wen2015improved}, the beta kernel estimator \citep{charpentier2007estimation}, and the mirror-reflection estimator \citep{gijbels1990estimating}. Among the above-mentioned kernel estimators, the transformation local likelihood estimator \citep{geenens2017probit} was found by \citet{nagler2017nonparametric} to have an overall best performance. The estimator is implemented in the R packages \texttt{kdecopula} \citep{kdecopula} and \texttt{rvinecopulib} \citep{rvinecopulib} using Gaussian kernels.
We  review its construction in Appendix~\ref{section:appendA}. To satisfy the copula definition, it is scaled to have uniform margins.

\noindent  As mentioned above  the simplifying assumption implies that 
$h_{U_i|U_j;\bms{U}_D}(u_i|u_j; \bm{u}_D)$ is independent of specific values of $\bm{u}_D$.
Thus it is sufficient to show how the h-function $h_{U_i|U_j}=C_{U_i|U_j}(u_i | u_j)$  can be estimated nonparametrically. For this we use as estimator
\begin{equation*}
	\hat{C}_{U_i|U_j}(u_i | u_j) = \int^{u_i}_0 \hat{c}_{U_i, U_j}(\tilde{u}_i, u_j) \mathrm{d} \tilde{u}_i
\end{equation*}
where $\hat{c}_{U_i, U_j}$ is  one of the above mentioned nonparametric estimators of the bivariate copula density of $(U_i, U_j)$ for which it holds that $\hat{c}_{U_i, U_j}$ integrates to 1 and has uniform margins.\\

\section{Vine based quantile regression}\label{section:vinebasedintro}

In the general regression framework the predictive ability of a set of variables $\bm{X} = (X_1, \ldots, X_p)^T$ for the response $Y\in\mathbbm{R}$ is studied. The main interest of vine based quantile regression is to predict the $\alpha \in (0, 1)$ quantile $
q_{\alpha}(x_1, \ldots, x_p) = F^{-1}_{Y|X_1, \ldots, X_p} (\alpha | x_1, \ldots, x_p)
$ of the response variable $Y$ given $\bm{X}$ by using  a copula based model of $(Y, \bm{X})^T$.
As shown in \citet{kraus2017d} this can be expressed as
\begin{equation}\label{eq:conditional quantile}
	F^{-1}_{Y|X_1, \ldots, X_p} (\alpha | x_1, \ldots, x_p)
	= F^{-1}_Y\big(C^{-1}_{V|U_1, \ldots,U_p}(\alpha | F_{X_1}(x_1), \ldots, F_{X_p}(x_p))\big),
\end{equation}
where $C_{V\vert U_1,\hdots , U_p}$ is the conditional distribution function of $V=F_Y(Y)$ given $U_j= F_{X_j}(X_j)= u_j$ for $j = 1,\hdots ,p$ with corresponding density $c_{V\vert U_1,\hdots , U_p}$, and $C_{V, U_1,\hdots , U_p}$ denotes the $(p+1)$-dimensional copula associated with the joint distribution of $(Y,\bm{X})^T$. In view of Section~\ref{introduction1}, we have $d= p+1$.
An estimate of $q_{\alpha}(x_1, \ldots, x_p)$  can be obtained using estimated marginal quantile functions $\hat{F}^{-1}_Y$, $\hat{F}^{-1}_{X_j}, j = 1, \ldots, p$ and the estimated conditional distribution function $\hat{C}^{-1}_{V |U_ 1, \ldots,U_ p}$ giving $\hat{q}_{\alpha}(x_1, \hdots, x_p) =\hat{F}^{-1}_Y \big( \hat{C}^{-1}_{V\vert U_1,\hdots , U_p}(\alpha\vert \hat{F}_{X_1}(x_1), \hdots \hat{F}_{X_p}(x_p))\big)$.

\noindent 	In general
$C_{V,U_1,\hdots ,U_p}$ can be any $(p+1)$-dimensional multivariate copula, however
for certain vine structures the corresponding conditional distribution function $C_{V\vert U_1,\hdots , U_p}$ can be obtained in closed form not requiring numerical integration. For D-vine structures this is possible and has been already utilized in \citet{kraus2017d}. \citet{masterMarija} showed that this is also the case for certain C-vine structures.  More precisely
the copula $C_{V, U_1,\hdots , U_p}$ with D-vine structure allows to express $C_{V|U_1,\hdots , U_p}$  in a closed form
if and only if the response $V$ is a leaf node in the first tree of the tree sequence. For a C-vine structure we need, that the node containing the response variable $V$ in the conditioned set is not a root node in any tree.
Additional flexibility in using such D- and C-vine structures is achieved by allowing for nonparametric pair-copulas as building blocks.

\noindent  The order of the predictors within the tree sequences itself is a free parameter with direct impact on the target function $C_{V\vert U_1,\hdots , U_p}$ and thus, on the corresponding prediction performance of $q_{\alpha}(x_1, \ldots, x_p)$.
For this we recall the concept of a node order for  C- and D-vine copulas introduced in \citet{masterMarija}. A D-vine copula denoted by $\mathcal{C}_D$ has order $ \mathcal{O}_{D}(\mathcal{C}_D)= (V,U_{i_1},\hdots ,U_{i_p}),$ if the response $V$ is the first node of the first tree $T_1$ and $U_{i_k}$ is the $(k+1)$-th node of $T_1$, for $k=1,\hdots ,p$. A C-vine copula  $\mathcal{C}_C$ has order  $\mathcal{O}_{C}(\mathcal{C}_C) = (V,U_{i_1},\hdots ,U_{i_p}),$ if $U_{i_1}$ is the root node in the first tree $T_1$, $U_{i_2}U_{i_1}$ is the root node in the second tree $T_2$, and $U_{i_k}U_{i_{k-1}}; U_{i_1},\hdots ,U_{i_{k-2}}$ is the root node in the $k$-th tree $T_k$ for $k=3, \hdots, p-1$.

\noindent Now our goal is to find an optimal order of D- or C-vine copula model with regard to a fit measure. This measure has to allow to quantify the explanatory power of a model. One such measure is the estimated conditional copula log-likelihood function as a fit measure. For $N$ i.i.d. observations $\bm{v} \coloneqq (v^{(1)},\hdots ,v^{(N)})^T\;\textrm{and}\; \bm{u}_j\coloneqq( u_j^{(1)},\hdots ,u_j^{(N)})^T,\; \textrm{for}\; j=1,\hdots ,p$ of the random vector $(V,U_1,\hdots ,U_p)^T$
we fit a C- or D-vine copula with order $\mathcal{O}(\hat{\mathcal{C}})=(V,U_1,\hdots ,U_p)$ using nonparametric pair copulas. We denote this copula by $\hat{\mathcal{C}}$, then the fitted conditional log-likelihood can be determined as
\begin{align*}
	cll & (\hat{\mathcal{C}},\bm{v}, (\bm{u}_1,\hdots ,\bm{u}_p)) = \sum_{n=1}^N \ln \hat{c}_{V\vert U_1,\hdots , U_p}(v^{(n)}\vert u_1^{(n)},\hdots ,u_p^{(n)}) = 
	\sum_{n=1}^N\Big[\ln \hat{c}_{V,U_1}(v^{(n)},u_1^{(n)})  + \\
	& \sum_{j=2}^{p}\ln \hat{c}_{V,U_j\vert U_1,\hdots ,U_{j-1}}(\hat{C}_{V\vert U_1,\hdots ,U_{j-1}}( v^{(n)}\vert u_1^{(n)},\hdots ,u_{j-1}^{(n)}),
	\hat{C}_{U_j\vert U_1,\hdots ,U_{j-1}}( u_j^{(n)}\vert u_1^{(n)},\hdots ,u_{j-1}^{(n)})) \Big].
\end{align*}
Penalizations for model complexity when parametric pair copulas are used can be added as shown in  \citet{masterMarija}. To define an appropriate penalty in the case of using nonparametric pair copulas is an open research question (see also Section \ref{section:disscussion}).

\section{Forward selection algorithms}\label{section:main}
Having a set of $p$ predictors, there are $p!$ different orders that uniquely determine  $p!$ C-vines and  $p!$ D-vines. Fitting and comparing  all of them is computationally inefficient. Thus, the idea is to have an algorithm that will sequentially choose the elements of the order, so that at every step the resulting model for the prediction of the conditional quantiles has the highest conditional log-likelihood.
Building upon the idea of \citet{kraus2017d} for the one-step ahead D-vine regression, we propose an algorithm which allows for more flexibility and which is less greedy, given the intention to obtain a globally optimal C- or D-vine fit. The algorithm builds the C- or D-vine step by step, starting with an order consisting of only the response variable $V$. Each step adds one of the predictors to the order based on the improvement of the conditional log-likelihood, while taking into account the possibility of future improvement, i.e. extending our view two steps ahead in the order. As  discussed in Section~\ref{sectionnonpar}, the pair copulas at each step are estimated nonparametrically in contrast to the parametric approach of \citet{kraus2017d}.
We present the implementation for both C-vine and D-vine based quantile regression in a single algorithm, in which the user decides whether to fit a C-vine or D-vine model based on the background knowledge of dependency structures in the data.
Implementation for a large data set is computationally challenging; therefore, randomization is introduced  to guarantee computational efficiency  in high dimensions.

\subsection{Two-step ahead forward selection algorithm for C- and D-vine based quantile regression}\label{twostep}

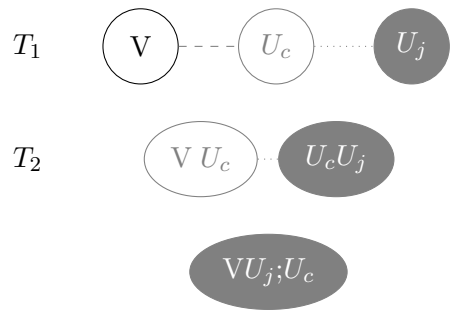
\begin{figure}[ht!]
	\centering
	\renewcommand{\xshiftNodes}{0.02*\linewidth}
	\renewcommand{\yshiftLabels}{.0cm}
	\begin{tikzpicture}	[every node/.style = VineNode, node distance =1.5cm]
		\node[TreeLabels] (T1)       {$T_1$}
		node             (v)         [right of = T1] {V}			
		node             (c)         [right of = v, xshift = \xshiftNodes,draw = gray, text=gray] {$U_c$}
		node             (j)         [right of = c, xshift = \xshiftNodes, draw = gray, fill= gray, text=white] {$U_j$}
		;
		\draw[dashed, color=gray] (v) to node[draw=none, fill = none, above, yshift = \yshiftLabels] {} (c);	
		\draw[dotted,color=gray] (c) to node[draw=none, fill = none, above, yshift = \yshiftLabels] {} (j);
		
		\node[TreeLabels] (T2)      [below of = T1] {$T_2$}
		node             (vc)         [right of = T2, xshift =0.8cm,draw = gray, text=gray] {V$\;U_c$}			
		node             (cj)         [right of = vc, xshift = \xshiftNodes,draw = gray, fill= gray, text=white] {$U_c U_j$}
		;
		\draw[dotted, color=gray] (vc) to node[draw=none, fill = none, above, yshift = \yshiftLabels] {} (cj);
		
		\node[DummyNode] (T3)      [below of = T2] {}
		node             (vcj)         [right of = T3, xshift =1.7cm, draw = gray, fill= gray, text=white] {V$U_j$;$U_c$};	
	\end{tikzpicture}
	\caption{ $V$ is fixed as the first node of $T_1$ and the first candidate predictor to be included in the model, $U_c$ (gray), is chosen based on the conditional log-likelihood of the two-step ahead copula including the predictor $U_j$ (gray filled).}
	\label{1stepDvine}
\end{figure}

\textbf{Input and data preprocessing:}
Consider $N$ i.i.d observations
$
\bm{y} \coloneqq (y^{(1)},\hdots ,y^{(N)})$ and $\bm{x}_j\coloneqq( x_j^{(1)},\hdots ,x_j^{(N)})\;\; \textrm{for}\; j=1,\hdots ,p ,
$
from the random vector $(Y,X_1,\hdots ,X_p)^T$. The  input data is on the x-scale, but in order to fit bivariate copulas we need to transform it to the u-scale using the probability integral transform. Since the marginal distributions are unknown we estimate them, i.e.  $F_{Y}$ and $F_{X_j}$, for $j=1,\hdots ,p,$ are estimated using a univariate nonparametric kernel density estimator with the R package \texttt{kde1d} \citep{kde1d}. This results in the pseudo copula data
$\hat{v}^{(n)} \coloneqq \hat{F}_Y(y^{(n)})\;$ and $\hat{u}_j^{(n)}\coloneqq\hat{F}_{X_j}(x_j^{(n)}),$ for $n=1,\hdots ,N,\;\; j=1,\hdots ,p.
$
The normalized marginals (z-scale) are defined as $Z_j\coloneqq\Phi^{-1}(U_j)$ for $j=1,\hdots ,p,$ and $Z_V\coloneqq\Phi^{-1}(V)$, where  $\Phi$ denotes the standard normal distribution function.
\\
\textbf{Step 1:}
To reduce computational complexity, we perform a pre-selection of the predictors based on Kendall's $\tau$. This is motivated by the fact that Kendall's $\tau$ is rank-based, therefore invariant with respect to monotone transformations of the marginals and can be expressed in terms of pair copulas. Using the pseudo copula data
$(\hat{\bm v}, \hat{\bm u}_j) = \lbrace \hat{v}^{(n)}, \hat{u}^{(n)}_j \vert n = 1,\hdots , N \rbrace,  $
estimates $\hat{\tau}_{VU_j}$ of the  Kendall's $\tau$  values  between the response $V$, and all possible predictors $U_j$ for $j=1,\hdots ,p$, are obtained.
For a given $k\leq p$, the $k$ largest estimates of $\vert\hat{\tau}_{VU_j}\vert$ are selected and the corresponding indices $q_1,\hdots ,q_k$ are identified such that
$\vert\hat{\tau}_{VU_{q_1}}\vert \geq \vert\hat{\tau}_{VU_{q_2}}\vert \geq\hdots\geq \vert\hat{\tau}_{VU_{q_k}}\vert \geq \vert\hat{\tau}_{VU_{q_{k+1}}}\vert \geq\hdots\geq \vert\hat{\tau}_{VU_{q_p}}\vert.$
The parameter $k$ is a hyper-parameter and therefore subject to tuning. To obtain a parsimonious model, we suggest a $k$ corresponding to $5\%$ - $20\%$ of the total number of predictors. The $k$ candidate predictors and the corresponding candidate index set of step 1 are defined as $U_{q_1},\hdots , U_{q_k}$ and $K_1 = \left\lbrace q_1,\hdots ,q_k\right\rbrace$, respectively.
For all $ c \in K_1$ and $j\in \left\lbrace 1,\hdots ,p\right\rbrace \setminus \left\lbrace c\right\rbrace$ the candidate two-step ahead C- or D-vine copulas are defined as the 3-dimensional copulas $\mathcal{C}^1_{c,j}$ with order $\mathcal{O}(\mathcal{C}^1_{c,j}) = (V,U_c,U_j)$. The first predictor is added to the order based on the conditional log-likelihood of the candidate two-step ahead C- or D-vine copulas, $\mathcal{C}^1_{c,j}$, given as
\begin{small}

\begin{equation*}
	cll\left(\mathcal{C}^1_{c,j},\bm{\hat{v}},(\bm{\hat{u}}_c,\bm{\hat{u}}_j)\right) 
	= \sum_{n=1}^N \Big[\log \hat{c}_{V,U_c}(\hat{v}^{(n)},\hat{u}_c^{(n)}) 
	+ \log \hat{c}_{V,U_j\vert U_c}\big( \hat{h}_{V|U_c} ( \hat{v}^{(n)}\vert\hat{u}_c^{(n)}),\hat{h}_{U_j|U_c}( \hat{u}_j^{(n)}\vert\hat{u}_c^{(n)})\big)\Big].
\end{equation*}

\end{small}
\noindent For each candidate predictor $U_c$, the maximal two-step ahead conditional log-likelihood at step 1, $cll_c^1$, is defined as
$	cll_c^1 \coloneqq \max_{j\in \lbrace 1,\hdots ,p\rbrace \setminus \lbrace c\rbrace} cll\left(\mathcal{C}^1_{c,j},  \bm{\hat{v}},(\bm{\hat{u}}_c,\bm{\hat{u}}_j)\right),\;\forall c\in K_1. $
Finally, based on the maximal two-step ahead conditional log-likelihood at step 1, $cll_c^1$, the index $t_1$ is chosen as
$	t_1 \coloneqq \argmax_{c\in K_1}\; cll_c^1,$
and the corresponding candidate predictor $U_{t_1}$ is selected as the first predictor added to the order. An illustration of the vine tree structure of the candidate two-step ahead copulas $\mathcal{C}^1_{c,j}$, in the case of fitting a D-vine model, with order $\mathcal{O}_{D}(\mathcal{C}^1_{c,j}) = (V,U_c,U_j)$ is given in Figure~\ref{1stepDvine}. Finally, the current optimal fit after the first step is the C-vine or D-vine copula, $\mathcal{C}_1$ with order  $\mathcal{O}(\mathcal{C}_1) = (V,U_{t_1})$. 

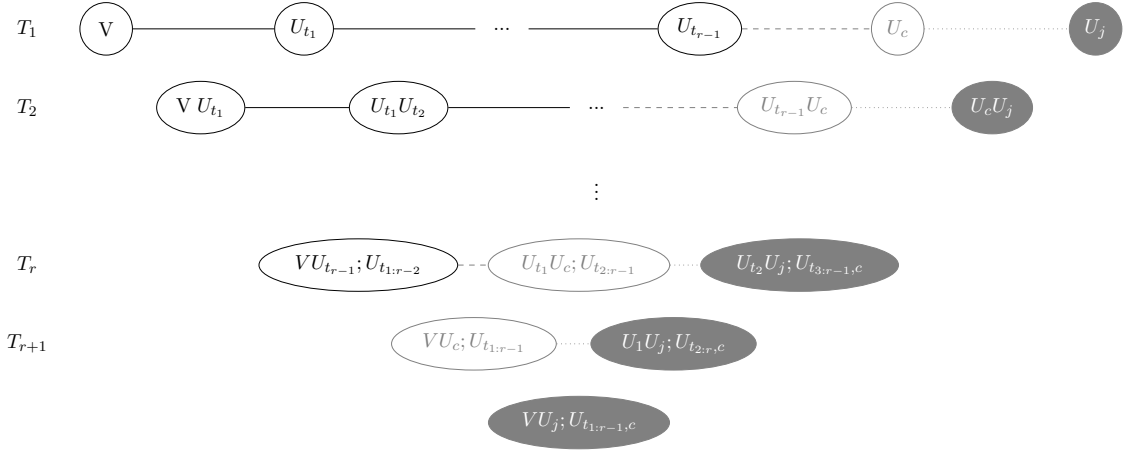
\begin{figure}[!h]
	\centering
	\renewcommand{\xshiftNodes}{0.15*\linewidth}
	\renewcommand{\yshiftLabels}{.0cm}
	\resizebox{6in}{!}{	
		\begin{tikzpicture}	[every node/.style = VineNode, node distance =1.5cm]
			
			\node[TreeLabels] (T1)       {$T_1$}
			node             (v)         [right of = T1] {V}			
			node             (u1)         [right of = v, xshift = \xshiftNodes] {$U_{t_1}$}
			node[DummyNode]             (u2)         [right of = u1, xshift = \xshiftNodes] {...}
			node             (u3)         [right of = u2, xshift = \xshiftNodes] {$U_{t_{r-1}}$}
			node             (c)         [right of = u3, xshift = \xshiftNodes,draw = gray, text=gray] {$U_c$}
			node             (j)         [right of = c, xshift = \xshiftNodes,draw = gray, fill= gray, text=white] {$U_j$}
			;
			\draw (v) to node[draw=none, fill = none, above, yshift = \yshiftLabels] {} (u1);
			\draw (u1) to node[draw=none, fill = none, above, yshift = \yshiftLabels] {} (u2);	
			\draw (u2) to node[draw=none, fill = none, above, yshift = \yshiftLabels] {} (u3);	
			\draw[dashed, color=gray] (u3) to node[draw=none, fill = none, above, yshift = \yshiftLabels] {} (c);
			\draw[dotted, color=gray] (c) to node[draw=none, fill = none, above, yshift = \yshiftLabels] {} (j);
			
			\node[TreeLabels] (T2)       [below of =T1]    {$T_2$}
			node             (1)         [right of = T2,xshift = 1.8cm] {V$\;U_{t_1}$}			
			node             (2)         [right of = 1, xshift = \xshiftNodes] {$U_{t_1}U_{t_2}$}
			node[DummyNode]            (3)         [right of = 2, xshift = \xshiftNodes] {...}
			node             (4)         [right of = 3, xshift = \xshiftNodes,draw = gray, text=gray] {$U_{t_{r-1}}U_c$}
			node             (5)         [right of = 4, xshift = \xshiftNodes, draw = gray, fill= gray, text=white] {$U_cU_j$}
			;
			\draw (1) to node[draw=none, fill = none, above, yshift = \yshiftLabels] {} (2);
			\draw (2) to node[draw=none, fill = none, above, yshift = \yshiftLabels] {} (3);	
			\draw[dashed, color=gray] (3) to node[draw=none, fill = none, above, yshift = \yshiftLabels] {} (4);
			\draw[dotted, color=gray] (4) to node[draw=none, fill = none, above, yshift = \yshiftLabels] {} (5);

			\node[TreeLabels] (T3)     [below of = T2]  {}
			node[DummyNode]  (Dummy3)     [below of = T2,    xshift = \xshiftNodes] {}	
			node[DummyNode]  (Dummy4)     [right of = Dummy3, xshift = \xshiftNodes] {}
			node[DummyNode]  (Dummy7)     [right of = Dummy4, xshift = 3.3cm] {$\vdots$}
			node[DummyNode]  (Dummy8)     [right of = Dummy7, xshift = \xshiftNodes] {}
			node[DummyNode]  (Dummy9)     [right of = Dummy8, xshift = \xshiftNodes] {}		;

			%
			
			\node[TreeLabels] (d)   [below of = T3]   {$T_r$}
			node      (10)  [right of = d, xshift= 4.8cm ]  {$VU_{t_{r-1}} ; U_{t_{1: r-2}}$}
			node      (11)  [right of = 10, xshift= 2.7cm ,draw = gray, text=gray]  {$U_{t_1}U_c;U_{t_{2:r-1}}$}
			node      (12)  [right of = 11, xshift= 2.7cm ,draw = gray, fill= gray, text=white]  {$U_{t_2}U_j;U_{t_{3:r-1},c}$}
			;
			\draw[dashed, color=gray] (10) to node[draw=none, fill = none, above, yshift = \yshiftLabels] {} (11);	
			\draw[dotted, color=gray] (11) to node[draw=none, fill = none, above, yshift = \yshiftLabels] {} (12);
			
			\node[TreeLabels] (T5)      [below of = d] {$T_{r+1}$}
			node             (13)         [right of = T5, xshift =7cm,draw = gray, text=gray] {$VU_c;U_{t_{1:r-1}}$} 		
			node             (14)         [right of = 13, xshift = 2.3cm,draw = gray, fill= gray, text=white] {$U_1U_j; U_{t_{2:r},c}$}	;
			
			\draw[dotted, color=gray] (13) to node[draw=none, fill = none, above, yshift = \yshiftLabels] {} (14);
			
			\node[DummyNode] (T6)      [below of = T5] {}
			node             (16)         [right of = T6, xshift =9cm,draw = gray, fill= gray, text=white] {$VU_j;U_{t_{1:r-1},c }$};
			
	\end{tikzpicture} }
	\caption{ In step $r$, the current optimal fit, $\mathcal{C}_{r-1}$ (black), is extended by one more predictor, $U_c$ (gray), to obtain the new current optimal fit $\mathcal{C}_r$ (black and gray), based on the conditional log-likelihood of the two-step ahead copula $\mathcal{C}^r_{c,j}$ which also includes the predictor $U_j$ (gray filled). (In the figure, we use the shortened notation $U_{t_{1:r-1}}$ instead of  writing $U_{t_{1}}, \ldots , U_{t_{r-1}}$  and we use $U_{t_{1:r-1},c}$  instead of $U_{t_{1}}, \ldots , U_{t_{r-1}}, U_c$.)}
	\label{rstepDvine}
\end{figure}

\noindent\textbf{Step \bm{r}:}
After $r-1$ steps, the current optimal fit is the C- or D-vine copula $\mathcal{C}_{r-1}$ with order $ \mathcal{O}(\mathcal{C}_{r-1}) = (V,U_{t_1},\hdots  ,U_{t_{r-1}})$. At each previous step $i$, the order of the current optimal fit is sequentially updated with the predictor $U_{t_i}$ for $i = 1,\hdots ,r-1$.
At the $r$-th step the next predictor candidate is to be included. To do so, the set of potential candidates is narrowed based on a partial correlation measure.
Defining a partial Kendall's $\tau$ is not straightforward and requires the notion of a partial copula, which is the average over the conditional copula given the values of the conditioning values (for example see \cite{gijbels2021study} and the references given there). In addition, the computation in the case of multivariate conditioning is very demanding and still an open research problem. Therefore we took a pragmatic view and base our candidate selection on partial correlation.
Due to the assumption of Gaussian margins inherited to the  Pearson's partial correlation, the estimates are computed on the z-scale.
Estimates of the empirical Pearson's partial correlation, $\hat{\rho}_{Z_V,Z_j;Z_{t_1},\hdots ,Z_{t_{r-1}}}$, between the normalized response variable $V$ and available predictors $U_j$ for $j\in\lbrace 1,2,\hdots ,p\rbrace\setminus\lbrace t_1,\hdots ,t_{r-1}\rbrace$ are obtained.
Similar to the first step, a set of candidate predictors of size $k$ is selected based on the largest values of $\vert\hat{\rho}_{Z_V,Z_j;Z_{t_1},\hdots ,Z_{t_{r-1}}}\vert$ and the corresponding indices $q_1,\hdots ,q_k$. 
The $k$ candidate predictors and the corresponding candidate index set of step $r$ are defined as $U_{q_1},\hdots , U_{q_k}$ and the set $K_r = \left\lbrace q_1,\hdots ,q_k\right\rbrace$, respectively.
For all $ c \in K_r$ and $j\in\left\lbrace 1,2,\hdots ,p\right\rbrace\setminus\left\lbrace t_1,\hdots ,t_{r-1},c\right\rbrace$ the candidate two-step ahead C- or D-vine copulas are defined as the copulas $\mathcal{C}^r_{c,j}$ with order $\mathcal{O}(\mathcal{C}^r_{c,j}) = ( V,U_{t_1},\hdots ,U_{t_{r-1}}, U_c, U_j)$.
There are $k(p-r)$ different candidate two-step ahead C- or D-vine copulas $\mathcal{C}^r_{c,j}$ (since we have $k$ candidates for the one-step ahead extension $U_c$, and for each,  $p-(r-1) -1$ two step ahead extensions $U_j$). Their corresponding conditional log-likelihood functions are given as
\begin{small}
\begin{equation*}
	\begin{split}
		c&ll\left(\mathcal{C}^r_{c,j},\right. \left. \bm{\hat{v}}, (\bm{\hat{u}}_{t_1}\hdots \bm{\hat{u}}_{t_{r-1}}, \bm{\hat{u}}_c,\bm{\hat{u}}_j)\right) = \; cll\left(\mathcal{C}_{r-1}, \bm{\hat{v}},(\bm{\hat{u}}_{t_1}\hdots \bm{\hat{u}}_{t_{r-1}} )\right)+ \\
		& \sum_{n=1}^N \log \hat{c}_{VU_{c};U_{t_1},\hdots ,U_{t_{r-1}}} \left( \hat{C}_{V\vert U_{t_1},\hdots ,U_{t_{r-1}}}\big( \hat{v}^{\left(n\right)}\vert \hat{u}_{t_1}^{\left(n\right)},\hdots ,\hat{u}_{t_{r-1}}^{(n)}\big), 
		\hat{C}_{U_{c}\vert U_{t_1},\hdots ,U_{t_{r-1}}}\big(\hat{u}_{c}^{(n)}\vert \hat{u}_{t_1}^{(n)},\hdots ,\hat{u}_{t_{r-1}}^{(n)}\big)  \right) \\
		& + \sum_{n=1}^N	\log \hat{c}_{VU_j;U_{t_1},\hdots ,U_{t_{r-1}},U_{c}} \left( \hat{C}_{V\vert U_{t_1},\hdots ,U_{t_{r-1}},U_{c}}\big( \hat{v}^{(n)}\vert \hat{u}_{t_1}^{(n)},\hdots ,\hat{u}_{t_{r-1}}^{(n)},\hat{u}_{c}^{(n)}\big),
		\right. \\
		& \left. \hskip 4.9cm
		\hat{C}_{U_j\vert U_{t_1},\hdots ,U_{t_{r-1}},U_{c}}\big(\hat{u}_{j}^{(n)}\vert \hat{u}_{t_1}^{(n)},\hdots ,\hat{u}_{t_{r-1}}^{(n)},\hat{u}_{c}^{(n)}\big)  \right).
	\end{split}
\end{equation*}
\end{small}
The $r$-th predictor is then added to the order based on the maximal two-step ahead conditional log-likelihood at Step~$r$, $cll_c^r$, defined as
\begin{equation}\label{cll_max}
	cll_c^r \coloneqq \max_{j\in\left\lbrace 1,2,\hdots ,p\right\rbrace\setminus\left\lbrace t_1,\hdots ,t_{r-1},c\right\rbrace} cll\left(\mathcal{C}^r_{c,j},\bm{\hat{v}},(\bm{\hat{u}}_{t_1}\hdots \bm{\hat{u}}_{t_{r-1}}, \bm{\hat{u}}_c, \bm{\hat{u}}_j )\right),\;\forall c\in K_r.
\end{equation}
The index $t_r$ is chosen as
$	t_r \coloneqq \argmax_{c\in K_r} \; cll_c^r,$
and the predictor $U_{t_r}$ is selected as the $r-$th predictor of the order. An illustration of the vine tree structure of the candidate two-step ahead copulas $\mathcal{C}^r_{c,j}$, for a D-vine model with order $\mathcal{O}_{D}(\mathcal{C}^r_{c,j}) = ( V,U_{t_1},\hdots ,U_{t_{r-1}}, U_c, U_j)$ is given in Figure~\ref{rstepDvine}. At this step, the current optimal fit is the C-vine or D-vine copula $\mathcal{C}_r$, with order
$\mathcal{O}(\mathcal{C}_r) = ( V,U_{t_1},\hdots U_{t_r}).$
The iterative procedure is repeated until all predictors are included in the order of the C- or D-vine copula model.

\subsubsection{Additional variable reduction in higher dimensions}\label{twostepred}

The above search procedure requires calculating $p-r$ conditional log-likelihoods for each candidate predictor at a given step $r$. This leads to calculating a total of $(p-r)k$ conditional log-likelihoods, where $k$ is the number of candidates. For $p$ large, this  procedure would cause a heavy computational burden. Hence, the idea is to reduce the number of conditional log-likelihoods calculated for each candidate predictor. This is achieved by reducing the size of the set, over which the maximal two-step ahead conditional log-likelihood $cll_c^r$ in \eqref{cll_max}, is computed. Instead of over the set $\left\lbrace 1,2,\hdots ,p\right\rbrace\setminus\left\lbrace t_1,\hdots ,t_{r-1},c\right\rbrace$, the maximum can be taken over an appropriate subset. This subset can be then chosen either based on the largest Pearson's partial correlations in absolute value denoted as $|\hat{\rho}_{Z_V,Z_j;Z_{t_1},\hdots ,Z_{t_{r-1}},Z_c}|$, by random selection, or a combination of the two. The selection method and the size of reduction are user-decided.

\subsection{Consistency of the conditional quantile estimator}
The conditional quantile function on the original scale in \eqref{eq:conditional quantile}
requires the inverse of the marginal distribution function of $Y$. Following \citet{kraus2017d, noh2013copula},
the marginal cumulative distribution functions $F_Y$ and $F_{X_j}, j=1,\ldots p$, are estimated nonparametrically to reduce the bias caused by model misspecification. Examples of nonparametric estimators for the marginal distributions $F_Y$ and $F_{X_j}$'s, are the continuous kernel smoothing estimator \citep{parzen1962estimation}
and the transformed local likelihood estimator in the univariate case \citep{geenens2014probit}.
Using a Gaussian kernel, the above two estimators of the marginal distribution are uniformly strong consistent. When also all inverses of the h-functions are estimated nonparametrically, we establish the consistency of the conditional quantile estimator $\hat{F}^{-1}_{Y|X_1, \ldots, X_p}$ in Proposition~\ref{thm:consistency} for fixed variable orders. By showing the uniform consistency, Proposition~\ref{thm:consistency} gives an indication
on the performance of the conditional quantile estimator $\hat{F}^{-1}_{Y|X_1, \ldots, X_p}$ for fixed variable orders, while combining the consistent estimators of $F_Y$, $F_{X_j}$'s, and bivariate copula densities. Under the consistency guarantee, the numerical performance of $\hat{F}^{-1}_{Y|X_1, \ldots, X_p}$ investigated by extensive simulation studies is presented in Section~\ref{section:simulation}.
\begin{proposition}\label{thm:consistency}
	Let the inverse of the marginal distribution functions $F_Y$ and $F_{X_j}$ $j=1,\ldots,p$ be uniformly continuous and estimated nonparametrically, and let the inverse of the h-functions expressing the conditional quantile estimator $C^{-1}_{V|U_1, \ldots, U_p}$ be uniformly continuous and estimated nonparametrically in the interior of the support of bivariate copulas, i.e., $[\delta, 1-\delta]^2, \delta \to 0_+$.
	\begin{enumerate}
		\item[1.] If estimators of the inverse of marginal functions $\hat{F}^{-1}_Y$, $\hat{F}^{-1}_{X_j}$, $j=1,\ldots,p$, are uniformly strong consistent on the support $[\delta,1 -\delta], \delta \to 0_+$, and the estimators of the inverse of h-functions composing the conditional quantile estimator $C^{-1}_{V|U_1, \ldots, U_p}$ are uniformly strong consistent, then the estimator
		$\hat{F}^{-1}_{Y|X_1, \ldots, X_p} (\alpha | x_1, \ldots, x_p)
		$
		is also uniformly strong consistent.
		\item[2.] If estimators of the inverse of marginal functions $\hat{F}^{-1}_Y$, $\hat{F}^{-1}_{X_j}$, $j=1,\ldots,p$, are at least weak consistent, and the estimators of the inverse of h-functions are also at least weak consistent, then the estimator
		$\hat{F}^{-1}_{Y|X_1, \ldots, X_p} (\alpha | x_1, \ldots, x_p)
		$ is weak consistent.
	\end{enumerate}
\end{proposition}
\noindent For more details about uniform continuous functions see \citet[Section~5.4]{bartle2000introduction}, \citet[p.109,Def.~1]{kolmogorov1970introductory}. 	
For a definition of strong uniform consistency or convergence with probability one, see \citet{ryzin1969,silverman1978weak} and \citet[p.16]{durrett2010probability}, while  for a definition for weak consistency or convergence in probability, see \citet[p.53]{durrett2010probability}.
The strong uniform consistency result in Proposition 1 requires additionally that all estimators of
$\hat{F}^{-1}_Y$, $\hat{F}^{-1}_{X_j}$, for $j=1,\ldots p$, are strong uniformly  consistent on a truncated compact interval $[\delta, 1 - \delta], \delta \to 0_+$. Although not directly used in the proof of Proposition~\ref{thm:consistency} in Appendix~\ref{section:appendB}, the truncation is an essential condition for guaranteeing the strong uniform consistency of all estimators of the inverse of the marginal distributions (i.e. estimators of quantile functions), see \citet{cheng1995uniform, van1998bootstrapping, cheng1984almost}.

\section{Simulation study}\label{section:simulation}
The proposed two-step ahead forward selection algorithms for C- and D-vine based quantile regression, from Section~\ref{twostep}, are implemented in the statistical language R \citep{Rlanguage}. The D-vine one-step ahead algorithm is implemented in the R package \texttt{vinereg} \citep{vinereg}.
In the simulation study from \cite{kraus2017d}, it is shown that the D-vine one-step ahead forward selection algorithm performs better or similar, compared to other state of the art quantile methods, boosting additive quantile regression \citep{Koenker2005quantile, fenske2011identifying}, nonparametric quantile regression \citep{li2013optimal}, semi-parametric quantile regression \citep{noh2015semiparametric}, and  the linear quantile regression \citep{koenker1978regression}. Thus we use the one-step ahead algorithm as the benchmark competitive method in the simulation study.
We set up the following simulation settings given below.
Each setting is replicated for $R = 100$ times. In each simulation replication, we randomly generate $N_{\rm train}$ samples used for fitting the appropriate nonparametric vine based quantile regression models. Additionally, another $N_{\rm eval} = \frac{1}{2}N_{\rm train}$ samples for Settings (a) -- (f) and $N_{\rm eval} = N_{\rm train}$ for Settings (g), (h) are generated for predicting conditional quantiles from the models. Settings (a) -- (f) are designed to test quantile prediction accuracy of nonparametric C- or D-vine quantile regression in cases where $p \leq N$; hence, we set $N_{\rm train} = 1000 \mbox{ or }300$. Settings (g) and (h) test quantile prediction accuracy in cases where $p > N$; hence, we set $N_{\rm train} = 100$.

\begin{enumerate}
	\item[(a)] Simulation Setting M5 from \citet{kraus2017d}:\\
	$$Y = \sqrt{|2X_1 - X_2 + 0.5 |} + (-0.5X_3 + 1)(0.1 X_4^3) + \sigma\varepsilon,$$ with $\varepsilon \sim N(0, 1), \sigma \in \{0.1, 1\}$, $(X_1, X_2, X_3, X_4)^T \sim N_4(0, \Sigma)$, and the $(i,j)$th component of the covariance matrix given as $(\Sigma)_{i,j} = 0.5^{|i - j|}$.
	\item[(b)] $(Y, X_1, \ldots, X_5)^T$ follows a mixture of two 6-dimensional t copulas with degrees of freedom equal to 3 and mixture probabilities 0.3 and 0.7. Association matrices $R_1$, $R_2$ and marginal distributions are recorded in  Table~\ref{table:marginal b}.

	\begin{table}[!htpb]
		\centering
		\begin{tabular}{c c}
			$R_1=
			\begin{pmatrix}
				1 & 0.6 & 0.5 & 0.6 & 0.7 & 0.1 \\
				0.6 & 1 & 0.5 & 0.5 & 0.5 & 0.5 \\
				0.5 & 0.5 & 1 & 0.5 & 0.5 & 0.5 \\
				0.6 & 0.5 & 0.5 & 1 & 0.5 & 0.5 \\
				0.7 & 0.5 & 0.5 & 0.5 & 1 & 0.5 \\
				0.1 & 0.5 & 0.5 & 0.5 & 0.5 & 1
			\end{pmatrix} $	
			&
			$ R_2 =
			\begin{pmatrix}
				1 & -0.3 & -0.5 & -0.4 & -0.5 & -0.1 \\
				-0.3 & 1 & 0.5 & 0.5 & 0.5 & 0.5 \\
				-0.5 & 0.5 & 1 & 0.5 & 0.5 & 0.5  \\
				-0.4 & 0.5 & 0.5 & 1 & 0.5 & 0.5 \\
				-0.5 & 0.5 & 0.5 & 0.5 & 1 & 0.5 \\
				-0.1 & 0.5 & 0.5 & 0.5 & 0.5 & 1
			\end{pmatrix} $ \\
		\end{tabular}
		\newline
		\vspace{0.3cm}
		\newline
		\begin{tabular}{c c c c c c }
			$Y$ & $X_1$ & $X_2$ & $X_3$ & $X_4$ & $X_5$ \\ \hline
			$N(0, 1)$ & $t_4$ & $N(1, 4)$ & $t_4$ & $N(1, 4)$ & $t_4$ \\		
		\end{tabular}
		\caption{Association matrices of the multivariate t-copula and marginal distributions for Setting (b).}
		\label{table:marginal b}.
	\end{table}
	%
	\item[(c)] Linear and heteroscedastic \citep{chang2019prediction}:\\ $Y =5(X_1 +X_2 +X_3 +X_4)+10(U_1 +U_2 +U_3 +U_4)\varepsilon,$ where $(X_1, X_2, X_3, X_4)^T \sim N(0, \Sigma)$, $\Sigma_{i,j} = 0.5^{I\{i \neq j\}}$, $\varepsilon \sim N_4(0, 0.5)$, and $U_j,$ $j  = 1,\ldots, 4$ are obtained from the $X_j$'s by the probability integral transform.
	\item[(d)] Nonlinear and heteroscedastic \citep{chang2019prediction}: \\ $Y = U_1U_2e^{1.8U_3U_4} + 0.5(U_1 + U_2 + U_3 + U_4)\varepsilon ,$  where $U_j, j = 1,\ldots, 4$ are probability integral transformed from $N_4(0, \Sigma)$, $\Sigma_{i,j} = 0.5^{I\{i \neq j\}}$, and $\varepsilon \sim N(0,0.5)$.
	\item[(e)] R-vine copula \citep{czado2019analyzing}: $(V, U_1, \ldots, U_4)^T$ follows an R-vine distribution with pair copulas given in Table~\ref{table:Rvinesample}.
	\begin{table}[ht]
		\centering
		\begin{tabular}{|cc|rcl|ccc|}
			\hline
			Tree & Edge & Conditioned & ; & Conditioning & Family & Parameter & Kendall's $\tau$ \\
			\hline
			1 & 1 &   $U_1,U_3$ & ; & & Gumbel & 3.9 & 0.74\\
			1 & 2 &   $U_2,U_3$ & ; & & Gauss & 0.9 &0.71 \\
			1 & 3 &   $V_{\;}\;,U_3$ & ; & & Gauss & 0.5& 0.33 \\
			1 & 4 &   $V_{\;}\;,U_4$ &   ; &  & Clayton & 4.8 & 0.71\\
			\hline
			2 & 1 &   $V_{\;}\;,U_1$ & ; & $U_3$ & Gumbel(90) & 6.5 & -0.85 \\
			2 & 2 &   $V_{\;}\;,U_2$ & ; & $U_3$ & Gumbel(90) & 2.6 & -0.62 \\
			2 & 3 &   $U_3,U_4$ &   ; & $V$ & Gumbel & 1.9 & 0.48 \\
			\hline
			3 & 1 &   $U_1,U_2$ & ; & $V_{\;}\;,U_3$ & Clayton & 0.9 &0.31 \\
			3 & 2 &   $U_2,U_4$ &   ; & $V_{\;}\;,U_3$ & Clayton(90) &5.1 &-0.72 \\
			\hline			
			4 & 1 &   $U_1,U_4$ &   ; & $V_{\;}\;,U_2,U_3$ & Gauss & 0.2 &0.13 \\
			\hline			
		\end{tabular}
		\caption{Pair copulas of the R-vine $C_{V,U_1,U_2,U_3,U_4}$, with their family parameter and Kendall's $\tau$  for Setting (e).}
		\label{table:Rvinesample}	
	\end{table}
	\item[(f)] D-vine copula \citep{masterMarija}: $(V,U_1,\ldots, U_5)^T$ follows a D-vine distribution with pair copulas given in Table~\ref{table:Dvinesample}.
	\begin{table}[ht]
		\centering
		\begin{tabular}{|cc|rcl|ccc|}
			\hline
			Tree & Edge & Conditioned & ; & Conditioning & Family & Parameter & Kendall's $\tau$ \\
			\hline
			1 & 1 &   $V_{\;}\;,U_1$ &   ; &  & Clayton & 3.00 & 0.60\\
			1 & 2 &   $U_1,U_2$ & ; & & Joe & 8.77 & 0.80\\
			1 & 3 &   $U_2,U_3$ & ; & & Gumbel & 2.00 & 0.50\\
			1 & 4 &   $U_3,U_4$ & ; & & Gauss & 0.20 & 0.13\\
			1 & 5 &   $U_4,U_5$ & ; & & Indep. & 0.00 & 0.00\\
			\hline
			2 & 1 &   $V_{\;}\;,U_2$ &   ; & $U_1$ & Gumbel & 5.00 & 0.80\\
			2 & 2 &   $U_1,U_3$ & ; & $U_2$ & Frank & 9.44 &0.65 \\
			2 & 3 &   $U_2,U_4$ & ; & $U_3$ & Joe & 2.78 & 0.49\\
			2 & 4 &   $U_3,U_5$ & ; & $U_4$ & Gauss & 0.20 & 0.13 \\
			\hline
			3 & 1 &   $V_{\;}\;,U_3$ &   ; & $U_1,U_2$ & Joe & 3.83 & 0.60 \\
			3 & 2 &   $U_1,U_4$ & ; & $U_2,U_3$ & Frank & 6.73 & 0.55\\
			3 & 3 &   $U_2,U_5$ & ; & $U_3,U_4$ & Gauss & 0.29 & 0.19\\
			\hline			
			4 & 1 &   $V_{\;}\;,U_4$ &   ; & $U_1,U_2,U_3$ & Clayton & 2.00 &0.50\\
			4 & 2 &   $U_1,U_5$ & ; & $U_2,U_3,U_4$ & Gauss & 0.09 &0.06 \\
			\hline		
			5 & 1 &   $V_{\;}\;,U_5$ &   ; & $U_1,U_2,U_3,U_4$ & Indep. & 0.00  &0.00\\
			\hline		
		\end{tabular}
		\caption{Pair copulas of the D-vine $C_{V,U_1,U_2,U_3,U_4,U_5}$, with their family parameter and Kendall's $\tau$  for Setting (f).}
		\label{table:Dvinesample}	
	\end{table}
	\item[(g)] Similar to Setting (a),\\
	$$Y = \sqrt{|2X_1 - X_2 + 0.5 |} + (-0.5X_3 + 1)(0.1 X_4^3) +  (X_5, \ldots, X_{110})(0, \ldots, 0)^T + \sigma\varepsilon,$$
	where $(X_1, \ldots, X_{110})^T \sim N_{110}(0, \Sigma)$ with the $(i, j)$th component of the covariance matrix $(\Sigma)_{i, j} = 0.5^{|i - j|}$, $\varepsilon \sim N(0, 1)$, and $\sigma \in \{0.1, 1\}$ .
	
	\item[(h)] Similar to (g),\\
	$Y = (X_1^{3}, \ldots, X_{110}^{3}) \bm{\beta} + \varepsilon ,$ where
	$(X_1, \ldots, X_{10})^T \sim N_{10}(0, \Sigma_A)$ with the $(i, j)$th component of the covariance matrix $(\Sigma_A)_{i, j} = 0.8^{|i - j|}$,  $(X_{11}, \ldots, X_{110})^T \sim N_{100}(0, \Sigma_B)$ with  $(\Sigma_B)_{i, j} = 0.4^{|i - j|}$.
	The first 10 entries of $\bm\beta$ are a descending sequence between $(2, 1.1) $ with increment of $0.1$  respectively, and the rest are equal to 0. We assume  $\varepsilon \sim N(0, \sigma)$ and  $\sigma \in \{0.1, 1\}$.
\end{enumerate}
\noindent Since the true regression quantiles are difficult to obtain in most settings, we consider the averaged check loss \citep{kraus2017d, komunjer2013quantile} and the interval score \citep{chang2019prediction, gneiting2007strictly}, instead of the out-of-sample mean averaged square error in \citet{kraus2017d}, to evaluate the performance of the estimation methods.
For a chosen  $\alpha \in (0,1)$, the averaged check loss is defined as
\begin{equation}\label{eq:check loss}
	\widehat{\mbox{CL}}_\alpha = \frac{1}{R} \sum_{r = 1}^R \bigg\{
	\frac{1}{N_{\rm eval}} \sum_{n = 1}^{N_{\rm eval}}\Big\{
	\gamma_\alpha \left(Y_{r,n}^{\rm eval} - \hat{q}_{\alpha}(X_{r, n}^{\rm eval}) \right)
	\Big\}
	\bigg\},
\end{equation}
where $\gamma_\alpha$ is the check loss function.

\noindent The interval score, for the $(1 - \alpha) \times 100\%$ prediction interval, is defined as
\begin{eqnarray}\label{eq:interval score}
	\lefteqn{\widehat{\mbox{IS}}_\alpha =  \frac{1}{R} \sum_{r = 1}^R \bigg\{
		\frac{1}{N_{\rm eval}} \sum_{n = 1}^{N_{\rm eval}}\Big\{
		\big(\hat{q}_{\alpha/2}(X_{r, n}^{\rm eval}) - \hat{q}_{1 - \alpha/2}(X_{r, n}^{\rm eval})\big) } \\ \nonumber
	&&+ \frac{2}{\alpha} \big( \hat{q}_{1-\alpha/2}(X_{r, n}^{\rm eval})  - Y_{r, n}^{\rm eval}\big) I\{Y_{r, n}^{\rm eval} \leq \hat{q}_{1-\alpha/2}(X_{r, n}^{\rm eval})\}\\
	&&+ \frac{2}{\alpha}\big( Y_{r, n}^{\rm eval} - \hat{q}_{\alpha/2}(X_{r,n}^{\rm eval})\big) I\{Y_{r, n}^{\rm eval} > \hat{q}_{\alpha/2}(X_{r, n}^{\rm eval})\}
	\Big\}
	\bigg\}, \nonumber
\end{eqnarray}
and smaller interval scores are better.
\begin{table}[!htp]
	\centering
	\def\arraystretch{1.1}
	\begin{tabular}{|c|c|c|c|c|c||c|c|c|c|}
		\hline
		Setting &Model & $\widehat{\mbox{IS}}_{0.05}$  &
		$\widehat{\mbox{CL}}_{0.05}$ &
		$\widehat{\mbox{CL}}_{0.5}$ &  $\widehat{\mbox{CL}}_{0.95}$
		& $\widehat{\mbox{IS}}_{0.05}$  &
		$\widehat{\mbox{CL}}_{0.05}$ &
		$\widehat{\mbox{CL}}_{0.5}$ &  $\widehat{\mbox{CL}}_{0.95}$
		\\ \hline
		&& \multicolumn{4}{c||}{$N_{{\rm train}}=300$}
		& \multicolumn{4}{c|}{$N_{{\rm train}}=1000$}\\
		\cline{1-10}
		(a) & D-vine One-step   & 55.54 & 0.66 & 0.16 & 0.51 & 55.89 &  0.67 & 0.15 & 0.50  \\
		$\sigma = 0.1$ &
		D-vine Two-step   & 43.33 & 0.47 & \cellcolor{mygray}0.10 & 0.41&
		40.74 & 0.45 & \cellcolor{mygray}0.09 & \cellcolor{mygray}0.37  \\
		**& C-vine One-step   & 53.51 & 0.64 & 0.16 & 0.49 &
		54.52 & 0.66 & 0.15 & 0.49  \\
		& C-vine Two-step    & 		\cellcolor{mygray}42.01 & 		\cellcolor{mygray}0.45 &		\cellcolor{mygray}0.10 &		\cellcolor{mygray}0.40  &
		\cellcolor{mygray}40.04 & 		\cellcolor{mygray}0.44 &		\cellcolor{mygray}0.09 &		\cellcolor{mygray}0.37  \\			
		\cline{1-10}
		(a) &D-vine One-step   & 154.35 &  1.63 &  \cellcolor{mygray}0.45 &  1.62 &
		162.12  & 1.70 &  0.43 &  1.66  \\
		$\sigma = 1$ &D-vine Two-step   & 148.53  & 1.57 &  		\cellcolor{mygray}0.45 &  		\cellcolor{mygray}1.56 &
		\cellcolor{mygray}156.77 &		\cellcolor{mygray}1.63 &  		\cellcolor{mygray}0.42&  \cellcolor{mygray}1.62 \\
		**& C-vine One-step   & 151.60  & 1.61 &\cellcolor{mygray}  0.45  & 1.60   &
		160.78 &  1.68 &  0.43 &  1.65   \\
		& C-vine Two-step    & 		\cellcolor{mygray}148.41 & 		\cellcolor{mygray}1.56 & 		\cellcolor{mygray}0.45 &  \cellcolor{mygray} 1.56
		& 156.79 &  \cellcolor{mygray}1.63 &  		\cellcolor{mygray}0.42 &  		\cellcolor{mygray}1.62 \\
		\cline{1-10}
		(b) & 		D-vine One-step   &	\cellcolor{mygray}118.75 &  		\cellcolor{mygray}1.29 &  0.42 &  		\cellcolor{mygray}1.30 &
		125.33 &  1.37 & 		\cellcolor{mygray}0.40 &  \cellcolor{mygray}1.36  \\
		*&D-vine Two-step   & 119.10 &  1.30 &  0.42 & \cellcolor{mygray} 1.30 &
		125.24 & 		\cellcolor{mygray}1.36 & \cellcolor{mygray} 0.40 &  \cellcolor{mygray}1.36  \\
		& C-vine One-step   & 119.08 &  1.30 &  		\cellcolor{mygray}0.41 & \cellcolor{mygray} 1.30&
		\cellcolor{mygray}125.12 &  \cellcolor{mygray}1.36 &  		\cellcolor{mygray}0.40 &   		\cellcolor{mygray}1.36   \\
		& C-vine Two-step    & 118.90 &  1.30 &  0.42 & 		\cellcolor{mygray}1.30  & 125.30 &  \cellcolor{mygray}1.36 &  \cellcolor{mygray}0.40 &  \cellcolor{mygray}1.36 \\
		\cline{1-10}
		(c) &D-vine One-step   & 2908.90  & 30.54 & 		\cellcolor{mygray}8.55 & 30.42 &
		3064.78 &  31.69  &  		\cellcolor{mygray}8.15 &  31.47  \\
		**&D-vine Two-step   & 2853.52 &  30.21  &  8.70  & 29.95 &
		\cellcolor{mygray}3041.95  &  		\cellcolor{mygray}31.61  &  8.20 &  31.26\\
		& C-vine One-step   & 2859.23 &  30.24  &  8.59  & 29.95 &
		3046.52  & 31.64  &  8.18 &  31.25 \\
		& C-vine Two-step    & 		\cellcolor{mygray}2850.10 &  		\cellcolor{mygray}30.19  &  8.64  & 		\cellcolor{mygray}29.84 &
		3042.46  & 31.62  &  8.20 &  		\cellcolor{mygray}31.23 \\
		\cline{1-10}
		(d) &D-vine One-step   & 86.40 & 0.92 & \cellcolor{mygray}0.24 & 0.91 & 91.11 & \cellcolor{mygray}0.96 &\cellcolor{mygray} 0.22 & 0.95\\
		**&D-vine Two-step   & 83.54 & 		\cellcolor{mygray}0.90 & 		\cellcolor{mygray}0.24 & 0.88 & 89.56 & \cellcolor{mygray}0.96 & \cellcolor{mygray}0.22 &  		\cellcolor{mygray}0.92 \\
		& C-vine One-step   & 84.99 & 0.91 & \cellcolor{mygray}0.24 & 0.90 & 90.40 &\cellcolor{mygray} 0.96 & \cellcolor{mygray}0.22 & 0.94 \\
		& C-vine Two-step    & \cellcolor{mygray}83.33 & \cellcolor{mygray}0.90 &\cellcolor{mygray} 0.24 & 		\cellcolor{mygray}0.87 & \cellcolor{mygray} 89.47 &  \cellcolor{mygray}0.96 & \cellcolor{mygray}0.22 &  \cellcolor{mygray}0.92 \\
		\cline{1-10}
		(e) &D-vine One-step   & 10.59 & 0.11 & \cellcolor{mygray}0.03 & 0.11 &
		10.49 & 0.11 & 0.03 & 0.11 \\
		*&D-vine Two-step   & 10.32 & 		\cellcolor{mygray}0.10 & \cellcolor{mygray} 0.03 & 0.11 & 10.26 & 		\cellcolor{mygray}0.09 & 		\cellcolor{mygray}0.02 & 0.11 \\
		& C-vine One-step   & 		\cellcolor{mygray}10.23 & 0.11 & \cellcolor{mygray}0.03 &  		\cellcolor{mygray}0.10 & \cellcolor{mygray}10.02 & 0.10 &\cellcolor{mygray} 0.02 &  		\cellcolor{mygray}0.10 \\
		& C-vine Two-step    & 10.35 & \cellcolor{mygray} 0.10 & 		\cellcolor{mygray}0.03 & 0.11 &  10.33 &   0.10 & 		\cellcolor{mygray}0.02 & 0.11 \\
		\cline{1-10}
		(f) &D-vine One-step   & 13.79 & 0.16 & 0.04 & 0.14 &13.70 & 0.16 & 0.04 & 0.14\\
		**&D-vine Two-step   & 		\cellcolor{mygray}8.44 &		\cellcolor{mygray}0.09 &		\cellcolor{mygray}0.02 & 		\cellcolor{mygray}0.08 & \cellcolor{mygray}8.28 & 		\cellcolor{mygray}0.09 &		\cellcolor{mygray}0.02 & 		\cellcolor{mygray}0.08 \\
		& C-vine One-step   & 12.62 & 0.14 & 0.04 & 0.13 &
		12.23 & 0.13 & 0.04 & 0.13 \\
		& C-vine Two-step    & 9.09 & 0.10 &\cellcolor{mygray} 0.02 & 0.09  &8.93 & \cellcolor{mygray}0.09 &\cellcolor{mygray} 0.02 & \cellcolor{mygray}0.08 \\
		\hline 
	\end{tabular}
	\caption{Out-of-sample predictions $\widehat{\mbox{IS}}_{0.5}$, $\widehat{\mbox{CL}}_{0.05}$, $\widehat{\mbox{CL}}_{0.5}$, $\widehat{\mbox{CL}}_{0.95}$ for Settings (a) -- (f) with $N_{{\rm train}}=300$ and $N_{{\rm train}}=1000$. Lower values, indicating better performance, are highlighted in gray. With ** we denote the scenarios in which there is an  improvement through the second step and with * we denote scenarios in which the models perform similar.}
	\label{setting300}
\end{table}

\begin{table}[!h]
	\centering
	\def\arraystretch{1.1}
	\begin{tabular}{|c|c|c|c|c||c|c|c|c|}
		\hline
		Model & $\widehat{\mbox{IS}}_{0.05}$  & $\widehat{\mbox{CL}}_{0.05}$ &
		$\widehat{\mbox{CL}}_{0.5}$ &  $\widehat{\mbox{CL}}_{0.95}$  
		& $\widehat{\mbox{IS}}_{0.05}$  & $\widehat{\mbox{CL}}_{0.05}$ &
		$\widehat{\mbox{CL}}_{0.5}$ &  $\widehat{\mbox{CL}}_{0.95}$  \\
		\hline
		& \multicolumn{4}{c||}{(g), $\sigma = 0.1$ *} & \multicolumn{4}{c|}{(g), $\sigma = 1$ **}\\\cline{2-9}
		D-vine One-step  &\cellcolor{mygray}19.63 & 0.26 & \cellcolor{mygray}0.25 & \cellcolor{mygray}0.23
		& 53.38 & 0.69 & 0.67 & 0.65  
		\\
		D-vine Two-step  & 20.48 & 0.26 & 0.26 & 0.25 & \cellcolor{mygray}52.17 &  0.68 & \cellcolor{mygray}0.65 &\cellcolor{mygray}0.63 \\
		C-vine One-step  & 19.73 & \cellcolor{mygray}0.25 & \cellcolor{mygray}0.25 & 0.24
		& 53.62 & 0.69 & 0.67 & 0.65
		\\
		C-vine Two-step   & 19.79 & \cellcolor{mygray}0.25 & \cellcolor{mygray}0.25 & 0.25 
		& 52.35& \cellcolor{mygray}0.67 & \cellcolor{mygray}0.65 & 0.64\\
		\hline
		& \multicolumn{4}{c||}{(h), $\sigma = 0.1$ **} & \multicolumn{4}{c|}{(h), $\sigma = 1$ **}\\
		\cline{2-9}
		D-vine One-step  &558.36 & 6.92 & 6.98&  7.04 & 554.18 &  6.87 & 6.93 &  6.99  \\
		D-vine Two-step  & 529.51&  6.46 & 6.62&  6.78  & 531.30 &  6.64 & 6.64 & 6.64    \\
		C-vine One-step  &  514.08& 6.05 & 6.43 & 6.81  & 512.96 &  6.39 &  6.41 & 6.44\\
		C-vine Two-step   &  \cellcolor{mygray}479.66 & \cellcolor{mygray}5.87&  \cellcolor{mygray}6.00 & \cellcolor{mygray}6.12  &  \cellcolor{mygray}483.92  & \cellcolor{mygray}6.05  &\cellcolor{mygray}6.05 & \cellcolor{mygray}6.05 \\		
		\hline
	\end{tabular}
	\caption{Out-of-sample predictions $\widehat{\mbox{IS}}_{0.5}$, $\widehat{\mbox{CL}}_{0.05}$, $\widehat{\mbox{CL}}_{0.5}$, $\widehat{\mbox{CL}}_{0.95}$ for Settings (g) -- (h) with $N_{\rm train}=100$. Lower values, indicating better performance, are highlighted in gray. With ** we denote the scenarios in which there is an improvement through the second step and with * we denote scenarios in which the models perform similar.}
	\label{settingbig}
\end{table}	

For Settings (a) -- (f), the estimation procedure for the two-step ahead C- or D-vine quantile regression follows exactly Section~\ref{twostep} where the candidate sets at each step include all possible remaining predictors. The additional variable reduction described in Section~\ref{twostepred} is not applied; thus, we calculate all possible conditional log-likelihoods in each step. On the contrary, due to computational burden in Settings (g) and (h), we set the number of candidates to be $k=5$ and the additional variable reduction from Section~\ref{twostepred} is applied. The chosen subset contains 20\% of all possible choices, where 10\% are predictors having the highest Pearson's partial correlation with the response and the remaining 10\% are chosen randomly from the remaining predictors.
Performance of the C- and D-vine two-step ahead quantile regression is compared with the C- and D-vine one-step ahead quantile regression. The performance of the competitive methods, evaluated by the averaged check loss at 5\%, 50\%, 95\% quantile levels and interval score for the 95\% prediction interval, are recorded in Tables~\ref{setting300} and \ref{settingbig}. All densities are estimated nonparametrically for a fair comparison.
Table~\ref{setting300}  shows that the C- and D-vine two-step ahead regression models outperform the C- and D-vine one-step ahead regression models in five out of seven settings, except Settings (b) and (e), in which all models perform quite similarly to each other. Again, when comparing regression models within the same vine copula class, the C-vine two-step ahead regression models outperform the C-vine one-step ahead models in five out of seven settings. Similarly, the D-vine two-step ahead models outperform the D-vine one-step ahead models in six out of seven scenarios, except Setting (b) only. In scenarios where there is no significant improvement through the second step, both one-step and two-step ahead approaches perform very similar. All of that implies that the two-step ahead vine based quantile regression greatly improves the performance of the one-step ahead quantile regression. Table~\ref{settingbig} indicates that in the high-dimensional settings, where the two-step ahead quantile regression was used in combination with the additional variable selection from Section~\ref{twostepred}, in three out of four simulation settings, the two-step ahead models outperform the one-step ahead models. In Setting (g), we can see that all models show similar performance. In Setting (g) with  standard deviation $\sigma =0.1$, the D-vine one-step ahead model outperforms the other models, while in Setting (g) with $\sigma=1$, the D-vine two-step ahead model shows a better performance. In Setting (h), we see a significant improvement in the two-step ahead models compared to the one-step ahead models. For both $\sigma =0.1$ and $\sigma =1$, the best performing model is the C-vine two-step ahead model.
These results indicate that the newly proposed method improves the accuracy of the one-step ahead quantile regression in high dimensions, even with an attempt to ease the computational complexity of the two-step ahead model with a low number of candidates, compared to the number of predictors.


\noindent 	
The proposed two-step algorithms, as compared
to the one-step algorithms are computationally more intensive.
We present the averaged computation time over $R =100$ replications on 100 paralleled cores (Xeon Gold 6140 CPUSs@2.6 GHz) in Settings~(g), (h) where $p > N_{\rm train}$, for the one step ahead and the two-step ahead approach. The high-dimensional settings have similar computational times since the computational intensity depends on the number of pair copula estimations and the number of candidates, which are the same for Settings~(g), (h). Hence, we only report the averaged computational times for Settings~(g), (h). The average computation time in minutes for the one-step ahead (C- and D-vine) approach is 83.01, in contrast to 200.28 by the two-step ahead (C- and D-vine) approach.  With the variable reduction from Section~\ref{twostepred}, the two-step algorithms double the time consumption of the one-step algorithms in exchange for prediction accuracy.


\section{Real data examples}\label{section: real data}
We test the proposed methods on two real data sets, i.e., the  Concrete data set from \citet{yeh1998modeling} corresponding to $p \leq N$, and the Riboflavin data set from \citet{buhlmann2011statistics} corresponding to $p > N$. For both, performance of the four competitive algorithms is evaluated by the averaged check loss defined in \eqref{eq:check loss} at 5\%, 50\% and 95\% quantile levels, and the 95\% prediction interval score defined in \eqref{eq:interval score}, by randomly splitting the data set into training and evaluation sets 100 times.
\subsection{Concrete data set}
\noindent The Concrete data set was initially used in \citet{yeh1998modeling}, and is available at the UCI Machine Learning Repository \citep{UCIrep}.
The data set has in total 1030 samples. Our objective is quantile predictions of the concrete compressive strength, which is a highly nonlinear function of age and ingredients. The predictors are age (\verb|AgeDay|, counted in days) and 7 physical measurements of the concrete ingredients (given in kg in a $m^3$ mixture): cement (\verb|CementComp|), blast furnace slag (\verb|BlastFur|), fly ash (\verb|FlyAsh|), water (\verb|WaterComp|), superplastizer (\verb|Superplastizer|), coarse aggregate  (\verb|CoarseAggre|) and fine aggregate  (\verb|FineAggre|).  We randomly split the data set into a training set with 830 samples and an evaluation set with 200 samples; the random splitting is repeated 100 times.
Performance of the proposed C- and D-vine two-step ahead quantile regression, compared to the C- and D-vine one-step ahead quantile regression, is evaluated by several measurements reported in Table~\ref{concrete} after 100 repetitions of fitting the models.  It is not unexpected that the results of the four algorithms are more distinct than most simulation settings, given the small number of predictors. However, there is an improvement in the performance of the two-step ahead approach compared to the one-step ahead approach for both C- and D-vine based models. Also, the C-vine model seems more appropriate for modeling the dependency structure in the data set. Finally, out of all models, the C-vine two-step ahead algorithm is the best performing algorithm in terms of out-of-sample predictions $\widehat{\mbox{IS}}_{0.5}$, $\widehat{\mbox{CL}}_{0.05}$, $\widehat{\mbox{CL}}_{0.5}$, $\widehat{\mbox{CL}}_{0.95}$ on the  Concrete data set, as seen in Table~\ref{concrete} .

\begin{figure}[!h]
	\includegraphics[width=\textwidth, height= 5cm]{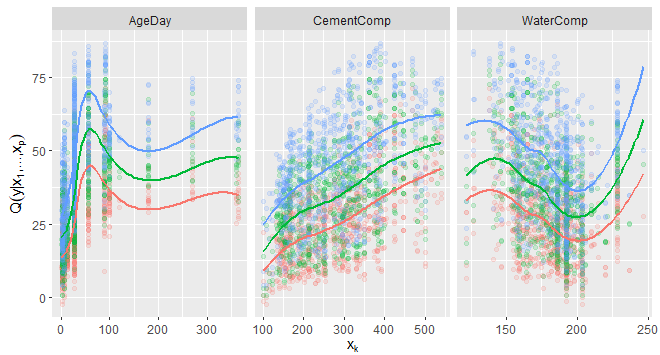}
	\centering
	\caption{Marginal effect plots for the 3 most influential predictors on the concrete compressive strength for $\alpha$ values of $ 0.05$ (red colour), $0.5$ (green colour) and $ 0.95$ (blue color).}
	\label{figure:quantconcrete}
\end{figure}

\begin{table}[!h]
	\centering
	\def\arraystretch{1.2}
	\begin{tabular}{|l|c||c|c|c|}
		\hline
		Model & $\widehat{\mbox{IS}}_{0.05}$  &
		$\widehat{\mbox{CL}}_{0.05}$ &
		$\widehat{\mbox{CL}}_{0.5}$ & $\widehat{\mbox{CL}}_{0.95}$ \\
		\hline
		D-vine One-step  & 1032.32 &  10.75 & 2.76 & 10.52  \\
		D-vine Two-step  & 987.10 &  10.54 & 2.78 & 9.82  \\
		C-vine One-step  & 976.75 &  10.65 & 2.70 & \cellcolor{mygray} 9.45 \\
		\rowcolor{mygray}
		C-vine Two-step  & 967.00 &  10.52 & 2.64 & 9.45 \\
		\hline
	\end{tabular}
	\caption{Concrete data set: Out-of-sample predictions $\widehat{\mbox{IS}}_{0.5}$, $\widehat{\mbox{CL}}_{0.05}$, $\widehat{\mbox{CL}}_{0.5}$, $\widehat{\mbox{CL}}_{0.95}$. The best performing model is highlighted in gray.}
	\label{concrete}
\end{table}
\noindent In Figure~\ref{figure:quantconcrete}  the marginal effect plots based on the fitted quantiles, from the C-vine two-step model,  for the  three most influential predictors are given. The marginal effect of a predictor is its expected impact on the quantile estimator, where the expectation is taken over  all other predictors. This is estimated using all fitted conditional quantiles and smoothed over the predictors considered.

\subsection{Riboflavin data set}
\begin{figure}[H]
	\includegraphics[width=\textwidth]{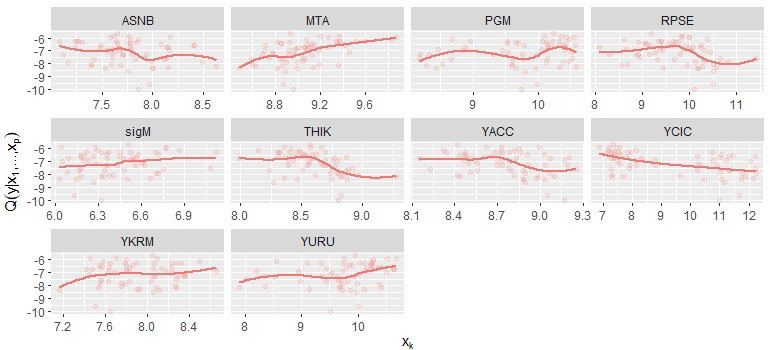}
	\centering
	\caption{Marginal effect plots for the 10 most influential predictors on the log-transformed Bacillus subtilis production rate for $\alpha = 0.5$.}
	\label{figure:ribo1}
\end{figure}

The Riboflavin data set, available in the R package \texttt{hdi}, aims at quantile predictions of the log-transformed production rate of Bacillus subtilis using log-transformed expression levels of 4088 genes. To reduce the computational burden, we perform a pre-selection of the top 100 genes with the highest variance \citep{buhlmann2011statistics}, resulting in a subset with $p = 100$ log-transformed gene expressions and $N =71$ samples. Random splitting of the subset into training set with 61 samples and evaluation set with 10 samples, is repeated for 100 times.
For the C- and D-vine two-step ahead quantile regression the number of candidates is set to $k=10$. Additionally, to further reduce the computational burden the additional variable selection from Section~\ref{twostepred} is applied with the chosen subset containing 25\% of all possible choices, where 15\% are predictors having the highest partial correlation with the log-transformed Bacillus subtilis production rate and the remaining 10\% are chosen randomly from the remaining predictors.
Performance of competitive quantile regression models is reported in Table~\ref{riboflavin}, where we see that the proposed C-vine two-step ahead quantile regression is the best performing model and  outperforms both the D-vine one-step ahead quantile regression from \citet{kraus2017d} and the C-vine one-step ahead quantile regression to a large extent. Further, the second best performing method is the D-vine two-step ahead model which, while performing slightly worse than the C-vine two-step ahead model, also significantly outperforms both the C-vine and D-vine one-step ahead models.
\begin{table}[!htp]
	\centering
	\def\arraystretch{1.2}
	\begin{tabular}{|c|c|c|c|c|}
		\hline
		Model & $\widehat{\mbox{IS}}_{0.05}$  &
		$\widehat{\mbox{CL}}_{0.05}$ &
		$\widehat{\mbox{CL}}_{0.5}$ &  $\widehat{\mbox{CL}}_{0.95}$ \\
		\hline
		D-vine One-step  & 33.83 &  0.44 &  0.42 & 0.41  \\
		D-vine Two-step  & 30.57 & 0.44 & 0.38 & 0.33 \\
		C-vine One-step  & 34.52 & 0.49 & 0.43 & 0.38 \\
		
		C-vine Two-step  & 		\cellcolor{mygray}28.59 & 		\cellcolor{mygray}0.41 & 		\cellcolor{mygray}0.36 & 		\cellcolor{mygray}0.30  \\
		\hline
	\end{tabular}
	\caption{Out-of-sample predictions $\widehat{\mbox{IS}}_{0.5}$, $\widehat{\mbox{CL}}_{0.05}$, $\widehat{\mbox{CL}}_{0.5}$, $\widehat{\mbox{CL}}_{0.95}$. The best performing model is highlighted in gray.}
	\label{riboflavin}
\end{table}
Since the predictors entering the C- and D-vine models yield a descending order of the predictors contributing to maximizing the conditional log-likelihood, the order indicates the influence  of the predictors to the response variable. It is often of practical interest to know which gene expressions are of the highest importance for prediction. Since we repeat the random splitting of the subset for $R = 100$ times, the importance of the gene expressions is ranked sequentially by choosing the one with the highest frequency of each element in the order excluding the gene expressions chosen in the previous steps. For instance, the most important gene expression is chosen as the one most frequently ranked first; the second most important gene is chosen as the one  most frequently chosen as the second element in the order, excluding the most important gene selected in the previous step. The top ten most influential gene expressions using the C- and D-vine one- or two-step ahead models are recorded in Table~\ref{table: top 10 gene serial numer}.
\begin{table}[!h]
	\centering
	\def\arraystretch{1.2}
	\begin{adjustbox}{max width=\textwidth,
			keepaspectratio}
		\begin{tabular}{|c|c|c|c|c|c|c|c|c|c|c|}
		  \hline
			Model/Position & 1  &  2&  3 &  4&  5 &  6 &  7 &  8&   9 &  10  \\ \hline
			D-vine One-step & GGT & YCIC & MTA & RPSE & YVAK & THIK & ANSB & SPOVB & YVZB & YQJB  \\
			D-vine Two-step & MTA & RPSE & THIK & YMFE & YCIC & sigM & PGM & YACC & YVQF & YKPB \\
			C-vine One-step & GGT & YCIC & MTA & RPSE & HIT & BFMBAB & PHRC & YBAE & PGM & YHEF  \\
			C-vine Two-step & MTA & RPSE & THIK & YCIC & YURU & PGM & sigM & YACC & YKRM & ASNB\\
			\hline
		\end{tabular}
	\end{adjustbox}
	\caption{The 10 most influential gene expressions on the conditional quantile function, ranked based on their position in the order.}
	\label{table: top 10 gene serial numer}
\end{table}
Figure~\ref{figure:ribo1}   shows the marginal effects plots based on the fitted quantiles, from the C-vine two-step model,  for the  10 most influential predictors on the log-transformed Bacillus subtilis production rate.

\section{Summary and discussion}\label{section:disscussion}
In this paper, we introduce a two-step ahead forward selection algorithm for nonparametric C- and D-vine copula based quantile regression. Inclusion of future information, obtained through considering the next tree in the two-step ahead algorithm, yields a significantly less greedy sequential selection procedure in comparison to the already existing one-step ahead algorithm for D-vine based quantile regression in  \citet{kraus2017d}.
We extend the vine-based quantile regression framework to include C-vine copulas, providing an additional choice for the dependence structure. Further, for the first time, nonparametric bivariate copulas are used to construct vine copula-based quantile regression models. The nonparametric estimation overcomes the problem of possible family misspecification in the parametric estimation of bivariate copulas and allows for even more flexibility in dependence estimation. Additionally, under mild regularity conditions, the nonparametric conditional quantile estimator is shown to be consistent.\\
\noindent The extensive simulation study, including several different settings and data sets with different dimensions, strengths of dependence and tail dependencies, shows that the two-step ahead algorithm outperforms the one-step ahead algorithm in most scenarios. The results for the Concrete and Riboflavin data sets are especially interesting, as the  C-vine two-step ahead algorithm has a significant improvement compared to the other algorithms. These findings provide strong evidence for the need of modeling the dependence structure following a C-vine copula. In addition,  the two-step ahead algorithm allows controlling the computational intensity independently of the data dimensions through the number of candidate predictors and the additional variable selection discussed in Section 5. Thus, fitting vine based quantile regression models in high dimensions becomes feasible. As seen in several simulation settings, there is a significant gain by
introducing additional dependence structures other then the D-vine based quantile regression.
A further research area is developing similar forward selection algorithms for R-vine tree structures while optimising the conditional log-likelihood.\\
\noindent At each step of the vine building stage, we compare equal-sized models with the same number of variables. The conditional log-likelihood is suited for such a comparison.
Other questions might come in handy, such as choosing between a C-vine, D-vine or R-vine information criteria. When maximum likelihood estimation is employed at all stages, the selection criteria by Akaike (AIC) \citep{Akaike73}, the Bayesian information criterion (BIC) \citep{Schwarz78} and the focussed information criterion (FIC) \citep{ClaeskensHjort2003} might be used immediately. \citet{KoHjortHobaekHaff2019} studied FIC and AIC specifically for the selection of parametric copulas. The copula information criterion in the spirit of the Akaike information criterion by \citet{GronnebergHjort2014} can be used for selection among copula models with empirically estimated margins, while \citet{KoHjort2019} studied such a criterion for parametric copula models. We plan a deeper investigation of the use of information criteria for nonparametrically estimated copulas and for vines in particular. Such a study is beyond the scope of this paper but could be interesting to study stopping criteria too for building vines.

\noindent Nonparametrically estimated vines are offering considerable flexibility. Their parametric counterparts, on the other hand, are enjoying simplicity. An interesting route for further research is to combine parametric and nonparametric components in the construction of the vines in an efficient way to bring the most benefit, which should be made tangible through some criterion such that guidance can be provided about which components should be modeled nonparametrically and which others are best modeled parametrically. For some types of models, such choice between a parametric and a nonparametric model has been investigated by \citet{JullumHjort2017} via the focussed information criterion. This and alternative methods taking the effective degrees of freedom into account are worth further investigating for vine copula models.



\section*{Acknowledgments}
We would like to thank the editor and the two  referees for their comments, which helped to improve the manuscript.
This work was supported by the Deutsche Forschungsgemeinschaft [DFG CZ 86/6-1],  the Research Foundation Flanders and KU Leuven internal fund C16/20/002.  The resources and services used in this work were provided by the VSC (Flemish Supercomputer Center), funded by the Research Foundation-Flanders (FWO) and the Flemish Government.	

\appendix

\section{Construction of the transformation local likelihood estimator of the copula density }\label{section:appendA}
Let the $N\times 2$ transformed sample matrix be
\begin{equation}\label{eq:transformed sample}
	D = (S, T),
\end{equation}
where the transformed samples $D_n = \big(S_n = \Phi^{-1}(U_i^{(n)}), T_n = \Phi^{-1}(U_j^{(n)})\big), n = 1, \ldots, N$, and $\Phi$ denotes the cumulative distribution function of a standard Gaussian distribution.
The logarithm of the density $f_{S, T}$ of the transformed samples $(S_n, T_n), n = 1, \ldots, N$ is approximated locally by a  bivariate  polynomial expansion $P_{\bms{a}_m}$ of order $m$ with intercept $\tilde a_{m,0}$ such that the approximation is denoted by
$$
\tilde f_{S, T}(\Phi^{-1}(u_i^{(n)}), \Phi^{-1}(u_j^{(n)})) =  \exp\big\{\tilde a_{m, 0}(\Phi^{-1}(u_i^{(n)}), \Phi^{-1}(u_j^{(n)})) \big\}.
$$
The transformation local likelihood estimator is then defined as
\begin{equation}\label{eq:kernel estimator}
	\tilde c (u_i^{(n)}, u_j^{(n)}) = \frac{\tilde f_{S, T}(\Phi^{-1}(u_i^{(n)}), \Phi^{-1}(u_j^{(n)}))}{\phi(\Phi^{-1}(u_i^{(n)})) \phi(\Phi^{-1}(u_j^{(n)}))}.
\end{equation}
To get the local polynomial approximation, we need a kernel function $\bm K$ with 2$\times$2 bandwidth matrix $\bm B_N$.
For some pair $(\check s, \check t)$ close to $(s, t)$, $\log f_{S T}(\check s, \check t)$ is assumed to be well approximated, locally, by for instance
a polynomial with $m = 1$ (log-linear)
\begin{eqnarray*}
	P_{\bms a_1}(\check s - s, \check t - t) =
	a_{1, 0}(s, t)
	+ a_{1, 1}(s, t) (\check s - s) +
	a_{1, 2}(s, t) (\check t - t),
\end{eqnarray*}
or $m = 2$ (log-quadratic)
\begin{eqnarray*}
	\lefteqn{P_{\bms a_2}(\check s - s, \check t - t)
		= a_{2, 0}(s, t)
		+ a_{2, 1}(s, t) (\check s - s) + a_{2, 2}(s, t) (\check t - t) } & \\
	&+  a_{2, 3}(s, t) (\check s - s)^2 +
	a_{2, 4}(s, t) (\check t - t)^2 +
	a_{2, 5}(s, t) (\check s - s)(\check t - t).
\end{eqnarray*}
The coefficient vector of the polynomial expansion $P_{a_m}$ is denoted by $a_m(s,t)$, where $a_1(s,t) = (a_{1,0}(s,t),\allowbreak a_{1,1}(s,t), a_{1,2}(s,t))$ for the log-linear approximation and $a_2(s,t) = (a_{2,0}(s,t),\ldots, a_{2,5}(s,t))$ for the log-quadratic. The estimated coefficient vector $\tilde{\bm a}_m(s,t)$ is obtained by a maximization problem in \eqref{eq:kernel polynomial term}
\begin{eqnarray}\label{eq:kernel polynomial term}
	\lefteqn{\tilde{\bm a}_m(s,t)
		= \arg\max_{a_m} \bigg\{ \sum_{n = 1}^N
		\bm{K} \bigg(\bm{B}_N^{-1/2}
		\begin{pmatrix}
			s - S_n \\
			t - T_n
		\end{pmatrix} \bigg) P_{\bms{a}_m}(S_n - s, T_n - t)}&&\nonumber\\
	&& -N \Big\{\int\!\!\int_{\mathbbm R^2}
	\bm{K} \bigg(\bm{B}_N^{-1/2}
	\begin{pmatrix}
		s - \check s \\
		t - \check t
	\end{pmatrix}
	\bigg) \exp\Big(P_{\bms{a}_m}(\check s  - s, t - t)\Big) \mathrm{d}\check s \mathrm{d} \check t
	\Big\}\bigg\}.
\end{eqnarray}
\noindent While it is well-known that kernel estimators suffer from the curse of dimensionality, in the vine construction only two-dimensional functions need to be estimated, this thus avoids problems with high-dimensionality.
\\
We next explain as in \citet{geenens2017probit} how a bandwidth selection is obtained.
Consider the principal component decomposition for the $N \times 2$ sample matrix $D = (S, T)$ in \eqref{eq:transformed sample}, such that the $N\times 2$ matrix $(Q, R)$ follows
\begin{equation}
	(Q, R)^T = {W} D^T,
\end{equation}
where each row of $W$ is an eigenvector of $D^T D$. We obtain an estimator of $f_{S T}$ through the density estimator of $f_{QR}$, which can be estimated based on a diagonal bandwidth matrix $\text{diag}(h_Q^2, h_R^2)$. Selecting the bandwidths $h_Q$ uses samples $Q_n, n = 1, \ldots, N$ as
\begin{eqnarray}\label{eq:bandwidth matrix}
	h_Q = \arg\min_{h > 0} \bigg\{\int_{-\infty}^{\infty} \Big\{\tilde f_Q^{(p)} \Big\}^2 dq - \frac{2}{N}\sum_{n = 1}^N\tilde f_{Q(-n)}^{(p)}(\hat Q_n)\bigg\},
\end{eqnarray}
where $\tilde f_Q^{(p)} (p  = 1, 2)$ are the local polynomial estimators for $f_Q$, and $\tilde f_{Q(-n)}^{(p)}$ is the ``leave-one-out" version of $\tilde f_{Q}^{(p)}$ computed by leaving out $Q_n$. The procedure of selecting $h_R$ is similar.
The bandwidth matrix for the bivariate copula density is then given by
$\bm{B}_N = K_N^{(p)} {W}^{-1} \textrm{diag}(h_Q^2, h_R^2) {W}^{-1}$ where $K_N^{(p)}$ takes
$N^{1/45}$ to ensure an asymptotic optimal bandwidth order for the local log-quadratic case ($p = 2$), see \citet[Section~4]{geenens2017probit} for details.
Selection for the k-nearest-neighbour type bandwidth is similar. The k-nearest-neighbour bandwidths denoted as $h'_Q$ and $h'_R$  are obtained by restricting the minimization in \eqref{eq:bandwidth matrix} in the interval $(0,1)$, i.e.,
$$
h'_Q = \arg\min_{h'_Q \in (0, 1)} \bigg\{\int_{-\infty}^{\infty} \Big\{\tilde f_Q^{(p)} \Big\}^2 dq - \frac{2}{N}\sum_{n = 1}^N\tilde f_{Q(-n)}^{(p)}(\hat Q_n)\bigg\}.
$$
Estimating $f_{QR}$ at any $(q, r)$ is obtained by using its $k = K_N^{(p)} \cdot h'_Q \cdot N$ nearest neighbours where $K_N^{(p)}$ takes
$N^{-4/45}$ for $p = 2$.  The R package \texttt{rvinecopulib} only implemented the bandwidth in \eqref{eq:bandwidth matrix} for the quadratic case with $p = 2$.

\section{Proof of Proposition~\ref{thm:consistency}} \label{section:appendB}
\begin{proof}
	We first show statement 1.
	By \eqref{eq:conditional quantile}, the estimator $\hat{F}^{-1}_{Y|X_1, \ldots, X_p} (\alpha | x_1, \ldots, x_p)
	=$\\ $ \hat{F}^{-1}_Y\big(\hat{C}^{-1}_{V|U_1, \ldots, U_p}(\alpha | \hat{u}_1, \ldots, \hat{u}_p)\big),$ where $\hat u_j = \hat F_j(x_j), j = 1, \ldots, p$ denote variables on the u-scale.
	To avoid heavy notation, $N$ referring to the sample size will be omitted here.
	Following \cite{wied2012consistency, silverman1978weak}, to show the uniformly strong consistency of $\hat{F}^{-1}_Y\big(\hat{C}^{-1}_{V|U_1, \ldots, U_p}(\alpha | \hat{u}_1, \ldots, \hat{u}_p)\big)$, we show $$\sup_\alpha \big|\hat{F}^{-1}_Y\big(\hat{C}^{-1}_{V|U_1, \ldots, U_p}(\alpha | \hat{u}_1, \ldots, \hat{u}_p)\big) - F^{-1}_Y\big(C^{-1}_{V|U_1, \ldots, U_p}(\alpha | u_1, \ldots, u_p)\big)  \big| \to 0 \ a.s. $$
	To improve the readability and simplify the notation in the proof, we first introduce some shorthand notation. Define
	$$D_{C,1} = \hat{F}^{-1}_Y\big(\hat{C}^{-1}_{V|U_1, \ldots, U_p}(\alpha | \hat{u}_1, \ldots, \hat{u}_p)\big), D_{C,2} = {F}^{-1}_Y\big(\hat{C}^{-1}_{V|U_1, \ldots, U_p}(\alpha | \hat{u}_1, \ldots, \hat{u}_p)\big),$$
	$$D_{C,3} =  \hat{F}^{-1}_Y\big(C^{-1}_{V|U_1, \ldots, U_p}(\alpha | u_1, \ldots, u_p)\big), D_{C,4} = {F}^{-1}_Y\big(C^{-1}_{V|U_1, \ldots, U_p}(\alpha | u_1, \ldots, u_p)\big),$$
	and the two differences $D_C = D_{C,1} - D_{C,3}$ and $D_F = D_{C, 3} - D_{C,4}.$
	\\
	For all $\epsilon \geq 0$,
	\begin{eqnarray}\label{eq:converge uniform}
		1  &\geq&
		P\big(\sup_\alpha \big|D_{C,1} - D_{C,4}  \big| \leq \epsilon \big) \nonumber 
		=P\big(\sup_\alpha \big| D_{C,1} - D_{C,3} + D_{C,3} - D_{C,4}  \big| \leq \epsilon  \big) \\
		& = & P\big(\sup_\alpha \big| D_C + D_F  \big| \leq \epsilon  \big)\nonumber\\
		&\geq&  P\big(\sup_\alpha \big\{\big|D_C \big| + \big|D_F  \big| \big\}  \leq \epsilon  \big) \geq  P\big(\sup_\alpha \big|D_C\big| + \sup_\alpha\big|D_F \big|   \leq \epsilon  \big)\nonumber \\
		&\geq&  P\big( \big(\sup_\alpha \big|D_C \big| \leq \frac{3}{4}\epsilon \big)
		\cap
		\big(\sup_\alpha\big|D_F\big|   \leq \frac{1}{4}\epsilon  \big) \big) \nonumber\\
		&=& P\Big( \big(\sup_\alpha \big|D_C\big| \leq \frac{3}{4}\epsilon\big) \;\big|
		\; \big(\sup_\alpha\big|D_F \big|  \leq \frac{1}{4}\epsilon \big)
		\Big)
		\cdot
		P\big(\sup_\alpha\big| D_F \big|  \leq \frac{1}{4}\epsilon \big).
	\end{eqnarray}
	Denote the event
	$A=\sup_\alpha\big|D_F \big|   \leq \frac{1}{4}\epsilon$,  $P(A) = 1$ holds by the uniform strong consistency of the estimator of $F_Y^{-1}$.
	Next,we show that the conditional probability in \eqref{eq:converge uniform} is equal to 1.
	\begin{eqnarray*}
		\lefteqn{P\big( (\sup_\alpha |D_C| \leq \frac{3}{4}\epsilon) \;\big| A\big)
			=  P\big( \sup_\alpha |D_{C,1} -
			D_{C,2} + D_{C,2} - D_{C,3}
			+ D_{C,4} - D_{C,4}| \leq \frac{3}{4}\epsilon \big| A 
			\big) }\\
		&\geq & P\big( \sup_\alpha |D_{C,1} - D_{C,2}|
		+ \sup_\alpha |D_{C,4} - D_{C,3}|
		+ \sup_\alpha |D_{C,2} - D_{C,4}| \leq \frac{3}{4}\epsilon \big| A 
		\big).
	\end{eqnarray*}
	This conditional probability is equal to 1, since the first and second supremum are less than or equal to $\frac{1}{4}\epsilon$ by conditioning on $A$ 
	and due to the uniform consistency of $\hat{F}^{-1}_Y$.
	The last supremum is less than or equal to $\frac{1}{4}\epsilon$ by \citet[][Thm.2]{bartle1961preservation} on almost uniform convergence, applied to the continuous inverse distribution function $F^{-1}_Y$, and taking the measurable space to be the probability space.
	First, $P\big( \sup_\alpha \big|\big(\hat{C}^{-1}_{V|U_1, \ldots, U_p}(\alpha | \hat{u}_1, \ldots, \hat{u}_p)\big) - \big(C^{-1}_{V|U_1, \ldots, U_p}(\alpha | u_1, \ldots, u_p)\big) \big| \leq \frac{1}{4}\epsilon \big) = 1$, which can be argued similar to \eqref{eq:converge uniform} using the uniform consistency and continuity of the inverse of the h-functions.
	Next, \eqref{eq:converge uniform} states
	$ P( \sup_\alpha |D_{C,1} - D_{C,4}| \leq \epsilon ) = 1.$
	We conclude that $\hat{F}^{-1}_Y\big(\hat{C}^{-1}_{V|U_1, \ldots, U_p}(\alpha | \hat{u}_1, \ldots, \hat{u}_p)\big)$ is uniformly strong consistent.
	\\
	To prove the weak consistency in 2, by \cite{wied2012consistency, silverman1978weak}, we only need to show
	$ P(|D_{C,1} - D_{C,4}| \leq \epsilon ) \to 1.$
	Using the same technique as in \eqref{eq:converge uniform} and a similar argument for proving statement 2 of Proposition~\ref{thm:consistency} with Theorem 2 on convergence in measure in \cite{bartle1961preservation}, the weak consistency can be obtained.
	
\end{proof}

\bibliographystyle{apalike}

\bibliography{CombinedReference}

\end{document}